\documentclass[12pt]{iopart}
\usepackage{amsmath}
\usepackage[latin1]{inputenc}
\usepackage[english]{babel}
\usepackage{latexsym}
\usepackage{graphicx}

\def\N{\mathcal{N}}   
\def\hN{\hat{\mathcal{N}}}
\def\be{\begin{equation}}
\def\ee{\end{equation}}
\def\bc{\begin{center}}
\def\ec{\end{center}}
\def\bea{\begin{eqnarray}}
\def\eea{\end{eqnarray}}
\def\bean{\begin{eqnarray*}}
\def\eean{\end{eqnarray*}}

\def\nn{\nonumber}
\def\ov{\overline}
\def\log{{\rm \; log}}
\def\det{{\rm \; det}}

\def\l{\lambda}
\def\a{\alpha}
\def\b{\beta}
\def\d{\delta}
\def\p{\partial}
\def\s{\sigma}
\def\hs{\hat{\sigma}}
\def\t{\tau}
\def\la{\langle}
\def\ra{\rangle}
\def\Eq{E_{eq}}
\begin{document}

\title[]{Spin-Glass Theory for Pedestrians}

\author{Tommaso Castellani$^\dagger$  and Andrea Cavagna$^\ddagger$}

\address{$\dagger$ Dipartimento di Fisica, Univesita' di Roma la Sapienza, Piazzale Aldo Moro 5, 00185 Roma

$\ddagger$ Istituto Sistemi Complessi, INFM - CNR, Via dei Taurini 19, 00185 Roma}

\begin{abstract}
In these notes the main theoretical concepts and techniques in the field of mean-field spin-glasses are reviewed in a
compact and pedagogical way, for the benefit of the graduate and undergraduate student. One particular spin-glass
model is analyzed (the $p$-spin spherical model)
by using three different approaches. Thermodynamics, covering pure states, overlaps, overlap distribution, 
replica symmetry breaking, and the static transition. Dynamics, covering the generating functional method, 
generalized Langevin equation, equations for the correlation and the response, the 
Mode Coupling approximation, and the dynamical transition. And finally complexity, covering the mean-field (TAP) free 
energy, metastable states, entropy crisis, threshold energy, and saddles. Particular attention has been 
paid on the mutual consistency of the results obtained from the different methods. 
\end{abstract}


\maketitle

\tableofcontents
\newpage
 \title[Spin-Glass Theory for Pedestrians]{}

\section{Introduction}

The aim of these notes is to provide graduate and undergraduate students in statistical physics with a sort of handbook 
of the main theoretical concepts in the physics of spin-glasses.  It is important to emphasize that this is not an 
overview of the entire field of disordered systems and spin-glasses: the whole experimental phenomenology is missing;
not a word is present on the large amount of numerical investigations and results; only one model is analyzed, compared to the 
vast number of different models on the spin-glass market; very little is said about  the connections
between spin-glasses and structural glasses (although {\it something} is said); and finally, the focus is entirely on mean-field
spin-glasses, leaving completely untouched what may be regarded as one of the most challenging open problems of the 
field, 
that is whether or not the mean-field picture has some validity also in finite dimensional systems. 

The student may thus rightfully ask {\it what} is contained in these notes. The basic idea is to present the most important 
theoretical techniques developed in the context of spin-glasses in a coherent, detailed, but at the same time very compact 
way. For this reason we study just one specific model, which we use as an ideal arena where to discuss, apply and compare
different theoretical methods. Although the model we consider has its own relevance in the field, the important point for us is to tell
the student a consistent and self-contained story, where each conceptual step has to be logically connected to the previous one.

In order to do this we had to necessarily disregard many important topics in the field, and at the same time to be 
very brief when introducing new ideas, hoping that  their practical implementation would help to
grasp their relevance. The perfect example is ergodicity breaking and pure states: an entire chapter, rather than few lines, 
should be devoted to these tricky, but crucial concepts. In this way, however, the notes would be unbearably long, and the main 
line of the story would quickly be lost. We opt for a synthetic exposure, leaving the student the freedom to go deeper on certain
subjects by a careful use of the extensive list of references.

The concepts and techniques developed in spin-glasses have found in recent years a wide range of applications in 
statistical physics and beyond, from biology to economics, passing through computer science and optimization theory. 
Our hope is that these notes may help the student to familiarize with the concepts, to practically learn how to handle 
them in a non-superficial way, and to eventually apply them to their own field of interest. The basic knowledge required
to follow these notes is just a reasonable preparation in standard statistical mechanics.

The three pillars of our discussion are Statics, Dynamics, and Complexity. 
The test system where all the calculations are done and the consistency of the different results is
analyzed, is the mean-field $p$-spin spherical model (PSM). This 
model is particularly apt to our purposes for more than one reason. 
First, the static (i.e. thermodynamic) analysis of the PSM gives results which are
drastically  different from the dynamical ones. For example, the two approaches give two different transition temperatures.
 This  naturally leads to the introduction of a third technique, dominated by the concept of complexity, which very nicely
 reconcile
static and dynamic results. Therefore, the PSM is the ideal model where to develop and compare the three approaches.
Secondly, the PSM is probably a simpler  model than the more famous and extensively studied  Sherrington-Kirkpatrick 
model, which after more than 25 year still puzzles us with its enormous variety of weird, yet very interesting, 
results. Finally, 
the PSM has some features which are intriguingly similar to structural glasses, most notably it is described by 
a set of dynamical equations which are identical to those provided by the Mode Coupling Theory for glasses. 
Therefore the PSM seems a good model to try and bridge the gap between spin-glasses and structural glasses.

As we have said, the target of these notes are graduate and undergraduate students. 
For this reason we tried to be as complete as possible when giving the details of the calculations, typically providing
more technical steps than it is usual in a technical paper. We hope that  in such a way it will always be possible for the
student to work out the final result. It is impossible to develop a 
genuine familiarity with spin-glasses without a serious training in the most technical aspects of the field.  We therefore encourage
the student to perform and check the calculations in these notes, in order to become as independent as possible when studying
similar subjects in her/his future. At the same time, we tried not to lose contact with the broader picture, and to always stick a sense
to any calculation we perform. In particular, we  stressed as much as possible the mutual consistency of results obtained 
with different techniques. The theory of spin-glasses is infamous for being  crowded with not-too-obvious formal steps, 
so it is always nice to  find the same result with two (or more) different, independent methods.

These notes are the expanded version of the lectures that one of us (AC) delivered in Bangalore, at the Conference and School on
Unifying Concepts in Glassy Physics (UCGP III), in June
2004, where also other lectures on different areas in the physics of glassy systems were presented. Wherever we could, we tried to
make contact, avoiding overlaps, with the notes of the other participants. 
In particular, we mention the Mode Coupling approximation in the section on Dynamics, in order to connect with the notes of
David Reichmann, and we restricted ourselves to equilibrium dynamics, given that the  subject of aging, and off-equilibrium 
dynamics in general, is extensively treated in the notes of Giulio Biroli.  We finally hope that the student will get the similarities 
between the chapter on the TAP approach and the energy landscape method analyzed by Francesco Sciortino.

We thank the organizers of UCGP III for giving us the opportunity to bring together in a single work what we hope will be
a useful collection of ideas and results in such a fascinating field of science.

\section{Basic concepts}

Before we start studying a specific spin-glass model, we need to introduce a couple of simple concepts and tricks, which we
will extensively use all along these notes. Each of them would deserve much more space than we can afford, and therefore
we  encourage the student to exploit the references. A background in statistical mechanics and in particular in the theory of
critical phenomena is very helpful. A nice and compact overview on this subject can be found in references \cite{cardy} and 
\cite{fisher-newman}. We also add here that a very nice and informal introduction to spin glasses can be found in \cite{hertz}, 
whereas the classic review for spin-glass theory is still reference \cite{me-pa-vi}.

\subsection{Disorder}

There are two main classes of disordered systems. The one spin-glasses belong to is that of {\it quenched disorder}.
In these systems the disorder is explicitly present in the Hamiltonian, typically under the form of
random couplings $J$ among the degrees of freedom $\sigma$,
\be
H=H(\s; J) \ .
\ee
The disorder $J$ is completely specified by its probability distribution $p(J) \, dJ$ which is the same for each 
different coupling constant in the system.  A famous  example is the Edwards-Anderson model \cite{edwards75},
\be
H= - \sum_{<ij>} J_{ij} \sigma_i \sigma_j \ ,
\ee
where the spins $\sigma_i=\pm1$ are the degrees of freedom, and the couplings $J_{ij}$ are
Gaussian random variables. This is a finite dimensional model, since the sum is performed over nearest-neighbor 
spins.  The disorder is {\it quenched}, meaning that the $J$ are constant on the time scale over which 
the $\sigma$ fluctuate. This will have a crucial consequence on the way we will have to perform the averages over $J$,
compared to $\sigma$. Spin-glasses are indeed systems with quenched  disorder.

Disorder creates frustration: it becomes impossible to satisfy {\it all} the couplings at the same time, as it would be
in a ferromagnetic system. Formally a system is frustrated if there exists a loop on which the product of the couplings is negative. 
In a frustrated loop, if
we fix an initial spin, and starting from it we try to chain-fix the other spins one after the other 
according to the sign of the 
couplings, we are bound to return to the initial spin and flip it. The only way to avoid frustration is 
to consider a lattice where there are no loops, for example a tree. Frustration is the main reason for the 
proliferation of metastable states in disordered systems.

In some system the disorder is not present in the Hamiltonian, but is in a way self-generated. This is the case of
structural glasses, whose Hamiltonian typically  takes the form,
\be
H= \sum_{ij} V(r_i- r_j)
\ee
where the degrees of freedom $r_i$ are the positions of the particles, and the function $V(r)$ is a 
deterministic potential (for example,  Lennard-Jones). Even though there is no quenched
disorder in the Hamiltonian, 
{\it at low temperature}, in a frozen glassy configuration of the system, each particle sees a different,
disordered  environment around itself. In this sense the disorder is self-generated. The origin of this phenomenon 
is the large number of non-crystalline local minima of the Hamiltonian.

It may seem odd that systems with quenched and self-generated disorder do have any property in common,
given their very different definitions. However, we shall see that {\it some} spin-glass models do have a 
phenomenology quite similar to the one of structural glasses.

\subsection{Self-averaging quantities}

In these notes we deal with spin-glasses, i.e. systems with quenched disorder in the Hamiltonian. Therefore, the
first key question is: How do we deal with the disorder?  The problem is that, in principle, each observable depends
on $J$, including the free energy of the system,
\be
F_N(J)=-\frac{1}{\b N}\log \int D\sigma e^{-\beta H(\sigma; J)}
\ee
where $N$ is the size of the system. This is very unpleasant, since it seems to suggest that the physical properties
of spin-glasses are different for each different realization of the disorder $J$, i.e. for each different 
sample. Were this true, it would be a disaster: we want to build a theory for spin-glasses, 
and not just for a specific piece of material ! In fact, both common sense and experience tells us that
for sufficiently large systems, physical properties {\it do not}  depend on $J$ anymore. Quantities like that are called
{\it self-averaging} \cite{me-pa-vi}, and the free energy is one of them,
\be
\lim_{N \to \infty}F_N(\b,J)=F_{\infty}(\b) \ .
\ee
In this case it is clear that the average over the disorder of a self-averaging quantity is equal to its $J$-independent value,
\be
F=-\lim_{N\to\infty}\frac{1}{\b N}\ov{\log Z(J)}=F_\infty(\b)
\ee
where,
\be
\ov{A}=\int dJ \, p(J)\,  A(J) \ .
\ee
This is good, since it means that analytically we can average over $J$, and that the result we obtain in this
way is in agreement with the physical value of the observable. Self-averageness is basically the same as asking
that the distribution of physical quantities is (for $N$ large) sharply peaked around their average value, that 
is that the variance of their distribution must go to zero for $N\to\infty$.
The free energy is self-averaging, and in particular,
\be
\ov{F^2}-{\ov{F}}^2=O\left(\frac1N\right) 
\label{lello}
\ee
If a quantity has, for example, a bimodal distribution, it is not self-averaging. Indeed its average is
a very poor indicator of the physical values of the quantity itself.

A simple argument to work out equation (\ref{lello})  can be given in finite dimension. We divide our system in a number $K$
of macroscopic sub-systems, with $1\ll K \ll N$. The total (extensive) free energy will be the sum of the free energies of the
sub-systems, plus a contribution coming from the interactions at the interfaces between the sub-systems. Once we compute 
the free energy {\it density}, this surface contribution
can be neglected in the limit $N\to\infty$. Moreover, the sub-systems free energies are independent random 
variables and therefore we can apply the
central limit theorem to the sum, and get (\ref{lello}).

\subsection{Annealed and quenched averages}
 
In order to average the free energy we have to compute the integral,
\be
F = -\frac{1}{\b N}\int dJ \; p(J)\; \log \int D\sigma \ e^{-\beta H(\sigma; J)} \ ,
\label{quenched}
\ee
which looks pretty bad, since we have to integrate a log over $J$. We could be tempted to 
define the following different quantity,
\be
F_a = -\frac{1}{\b N}\log \int dJ \; p(J)\; \int D\sigma \  e^{-\beta H(\sigma; J)} \ ,
\label{annealed}
\ee
which is certainly much simpler to compute. Unfortunately, this is not the right solution to our problem.
The difference between the two formulas above is in the role played by the disorder $J$: in (\ref{quenched})
we first integrate over the degrees of freedom, then take the log, and finally integrate over the random
couplings. In this way, the couplings $J$ are fixed, i.e. quenched, for each integration over the spins.
In other words, couplings and spins do not fluctuate together: for each realization of the disorder we
compute the free energy, and eventually we average it over $J$.
This kind of average is called {\it quenched}, and it is precisely what we need to do. 

On the other hand, it is clear that in (\ref{annealed}) the disorder $J$ and the degrees of freedom $\sigma$ have
been put on the same footing, fluctuating together. This is {\it not} what we want, since it means that
the time scale of variation of $J$ and $\sigma$ is the same, and therefore the disorder becomes yet another
degree of freedom, and it is no longer quenched. This second kind of average is called {\it annealed}, and, even
though it may be correct at high temperatures, where the frustration induced by the disorder is irrelevant, 
it is normally wrong at low temperatures, where the spins freeze in a state determined by the quenched value of
the couplings. A different way to see this point is that in the annealed case we are in fact averaging the
partition function $Z$, rather than the free energy $F$, over $J$. The fact is that $F$ is an extensive 
quantity, while $Z$ is not (it is {\it exponential} in $N$ !), and therefore $Z$ is not in general
self-averaging.

Therefore, we have to find a way to treat integrals like the one in (\ref{quenched}). This is where the
replica trick comes into play.

\subsection{The replica trick}

The replica trick \cite{edwards75} (which becomes in fact a method, when it is explained more deeply than here \cite{me-pa-vi}), 
stems from the following simple formula,
\be
\overline{\log \; Z} = \lim_{n\to 0} \frac1n \log \; \overline{Z^n} \ .
\ee
If $n$ remains a real number (as it should), there is no advantage at all in computing the r.h.s. compared to
the l.h.s., of course. However, if we now promote $n$ to be an {\it integer}, we can write:
\be
\overline{Z^n} = \int D\sigma_1 \dots D\sigma_n \ \overline{e^{-\b H(\sigma_1,J) \dots -\b H(\sigma_n,J)}}
\label{stropo}
\ee
which is in fact much simpler to compute. What we do is to replicate the system $n$ times, compute everything as
a function of $n$, and finally cross our finger in taking the limit $n\to 0$. It is crucial to understand that all the Hamiltonians
in (\ref{stropo}) have the {\it same} realization of the quenched disorder, and in this sense are replicas one of the other.

A different useful form of the replica trick is the following,
\bea
\overline{\langle A \rangle} = \overline{
\frac{1}{Z}\int D\sigma \; A(\sigma) \; e^{-\b H(\sigma,J)}} = 
\lim_{n\to 0} \overline{
Z^{n-1} 
\int D\sigma \; A(\sigma) \; e^{-\b H(\sigma,J)}}= \nn \\
\lim_{n\to 0} 
\int D\sigma_1 \dots D\sigma_n \ A(\sigma_1)\  \overline{e^{-\b H(\sigma_1,J) \dots -\b H(\sigma_n,J)}} \ .
\eea
Of course, the label $1$ we used for the replica into the observable $A$ is completely arbitrary, and thus we have to be careful
that our result {\it must not} depend on this particular index, otherwise we have a complete nonsense.

To conclude this small section, let us have a look to a case where the replica trick does work. Imagine that 
we ignore the rule $(x^a)^b=x^{ab}$ with $a,b$ real, but that we know that $x^m = x\cdot \dots \cdot x$, 
$m$ times. Given $y=x\cdot x$, we want to know what is $y^{1/2}$. We can use the replica trick:
\be
y^{1/2} = \lim_{n\to1/2} y^n = \lim_{n\to1/2} x\cdot x\cdot  \dots \cdot x\cdot x = \lim_{n\to1/2} x^{2n} = x\ .
\ee

\subsection{Pure states}

In the low temperature phase, and in the limit $N\to\infty$ we can have ergodicity breaking: the system 
at equilibrium explores only a sub-part of the phase space \cite{ruelle,parisi,me-pa-vi}. When this happens the Gibbs measure can
be split into sub-components, called {\it pure states},
\be
\langle\cdot\rangle = \sum_\alpha w_\alpha \langle\cdot\rangle_\alpha
\ee
where $\alpha$ is an index running over all the states, and $w_\alpha$ is the statistical (Gibbs) weight of 
state $\alpha$.
To better understand this formula, we must assume that it is possible to assign each configuration in the 
phase space with nonzero thermodynamics weight, to one and only one state. In this case we can write for any observable $A$,
\bea
\langle A \rangle =
\frac{1}{Z}\int D\sigma \ e^{\beta H(\Sigma)} A(\sigma) &=& 
\frac{1}{Z}\sum_\alpha \int_{\sigma\in\alpha}  D\sigma \ e^{\beta H(\Sigma)} A(\sigma) = \nn \\
\sum_{\alpha}\frac{Z_\alpha}{Z} \frac{1}{Z_\alpha} \int_{\sigma\in\alpha} D\sigma \ e^{\beta H(\Sigma)} A(\sigma) &=& 
\sum_{\alpha} w_\alpha \langle A \rangle_\alpha
\eea
where we have defined,
\be
Z_\alpha= \int_{\sigma\in\alpha} D\sigma \ e^{\beta H(\Sigma)}
\ee
that is the partition function restricted to state $\alpha$, and
\be
w_\alpha= \frac{Z_\alpha}{Z}
\ee
the statistical weight of state $\alpha$.

As an example we can consider the Ising model below $T_c$. In the thermodynamic limit the ergodicity is 
broken, and we have two states, with positive and negative spontaneous magnetization,
\be
\langle \cdot\rangle = \frac{1}{2} \langle\cdot\rangle_+ + \frac{1}{2} \langle\cdot\rangle_-
\ee
that is $w_+ = w_-= 1/2$, in absence of external magnetic field. It is crucial to split the measure,
otherwise we would not see any spontaneous magnetization,
\be
\langle \sigma \rangle  =  \frac{1}{2} \langle\sigma\rangle_+ + \frac{1}{2} \langle\sigma\rangle_- = 0
\ee
A very important feature of pure states is the {\it clustering} property. In essence, this property 
states the very physical concept that the statistical correlation between two different points goes
to zero when their distance goes to infinity, 
\be
\langle \sigma_i\sigma_j\rangle \to \langle \sigma_i\rangle\langle\sigma_j\rangle \quad \quad {\rm for}
\quad |i-j| \to \infty \ .
\ee
In other words, a very basic physical requirement is that connected correlation functions decay to 
zero at large distances \cite{parisi,me-pa-vi}. 
As we have said, this property only holds in pure states. Take, for example, the paramagnetic state
in the Ising model below $T_c$, that is the Gibbs ergodic measure over the full phase space:
\be
\langle \sigma_i\sigma_j\rangle = \frac{1}{2} \langle \sigma_i\sigma_j\rangle_+ +
 \frac{1}{2} \langle \sigma_i\sigma_j\rangle_- \to 
\frac{1}{2} \langle\sigma\rangle_+^2 + \frac{1}{2} \langle\sigma\rangle_-^2 = m^2 \neq 0 \ .
\ee
Therefore, the paramagnetic state is {\it not} a pure state below the critical temperature.

The example of the Ising model is particularly simple because we know {\it a priori} what  is 
the structure of pure states below $T_c$. In particular, we know how to select a state, i.e. how to project
the system onto any one of the two states: we simply apply a magnetic field. In disordered systems
the situation is not as simple as that, since {\it we do not know what is the field projecting the system
onto any particular state}. This crucial fact is at the heart of the difficulty in studying disordered systems:
we lack the magnetic field as a crucial tool to select states. Of course, given a state, there is a (disordered)
magnetic field selecting that state. The problem is that we do not know what this field is !

A final important remark. In finite-dimensional systems, only equilibrium states can break the
ergodicity, i.e. states with the lowest free energy {\it density}. In other words, the system cannot remain 
trapped for an infinite time in a metastable state, because in finite dimension free energy barriers
surrounding metastable states are always {\it finite}. The extra free energy of a droplet of size $r$ of 
equilibrium phase in a background metastable phase has a positive interface contribution 
which grows as $r^{d-1}$, and a negative volume contribution which grows as $r^d$,
\be
\Delta F = \sigma \, r^{d-1} - \delta f \, r^{d} \ , 
\ee
where here $\sigma$ is the surface tension and $\delta f$ is the bulk free energy difference between the two
phases. This function has always a maximum, whose finite height gives the free energy barrier to nucleation of the
equilibrium phase (note that at coexistence $\delta f=0$ and the barrier is infinite). 

Therefore, if initially in a metastable states
the system will, sooner or later, collapse in the stable state with lower free energy density. For this reason, in 
finite dimension we cannot decompose the Gibbs measure in metastable components. When this is done, it
is always understood that the decomposition is only valid for {\it finite} times, i.e times much smaller than the
time needed for the stable equilibrium state to take over.  On the other hand, in
mean-field systems (infinite dimension), barriers between metastable states may be infinite in the thermodynamic limit, 
and it is therefore possible to call 'pure states' also metastable states, and to assign them a Gibbs weight
$w_\alpha$.  We will analyze a mean-field spin-glass model, so that we will be allowed to perform the 
decomposition above even for  metastable states.

\subsection{Overlap, self-overlap}

In non-disordered magnetic systems, a good order parameter is normally the total average magnetization,
\be
m=\frac1 N \sum_{i=1}^N \la \s_i\ra
\ee
which is zero in the high temperature phase, and different from zero in the low temperature phase,
where the $\pm$ symmetry is broken. In disordered systems we may be tempted to use a similar order parameter,
\be
m=\frac1 N \sum_{i=1}^N \ov{\la \s_i\ra}
\ee
However, due to the disorder the local magnetizations in the low temperature phase are all frozen in different
directions (if the disorder distribution is unbiased, as we shall assume), and thus the magnetization defined
above is zero at all temperatures, even though the $\pm$ symmetry is physically broken for each spin 
in our sample.  A better order parameter is the Edward-Anderson parameter \cite{edwards75},
\be
q_{EA}=\frac 1 N\sum_{i=1}^N\ov{{\la \s_i\ra}^2}
\ee
Such a quantity is nonzero if the local magnetizations $m_i$ are locally nonzero, and thus is a good order parameter.
In fact $q_{EA}$ is a particular case of a more general quantity called {\it overlap}.

In our study of spin glasses we will often need a tool to measure the similarity of two configurations, or two 
states. To this aim we introduce the overlap. Given two configurations $\s$ and $\t$, we define their
mutual overlap as,
\be
q_{\s\t}=\frac 1 N\sum_{i=1}^N\s_i\t_i
\ee
With Ising spins $s_i=\pm 1$ we have that,
\be
q_{\s\t}=
\left\{
\begin{array}{rcl}
1 & \mbox{if} &  \mbox{$\s$ e $\t$ almost coincide} \\
-1 & \mbox{if} & \mbox{$\s$ e $\t$ are anti-correlated} \\
0 & \mbox{if}  & \mbox{$\s$ e $\t$ are totally uncorrelated} \\
\end{array}
\right.
\ee
The overlap is thus a measure of the similarity among different configurations. We can also compute the 
overlap of a configuration with itself, the {\it self-overlap},
\be
q_{\s\s}=\frac 1 N\sum_{i=1}^N\s_i\s_i
\ee
With Ising spins $q_{\s\s}=1$. In the following we will always deal with systems where the self-overlap 
of {\it configurations} is $1$.

The overlap can measure also the similarity between {\it states}: if the Gibbs measure is split into 
sub-components $\alpha$ due to ergodicity breaking, we define the overlap between states $\a$ and $\b$ as,
\be
q_{\a\b}=\frac1N\sum_{i=1}^N{\la\s_i\ra}_{\a}{\la\s_i\ra}_{\b}
\ee 
which can also be written as,
\bea
q_{\a\b}=\frac1N\sum_{i=1}^N \frac1Z_\a\int_{\s\in\a}D\s \ \s_i e^{-\b H(\s)}\frac1Z_\b\int_{\t\in\b} D\t\ \t_i e^{-\b H(\t)}=\nn \\
\frac{1}{Z_\a Z_\b}\int_{\s\in\a}\int_{\t\in\b}D\s\ D\t \  e^{-\b H(\s)} e^{-\b H(\t)} \ q_{\s\t}
\eea
This expression shows that by measuring the overlap among states, we are in fact measuring the overlaps among
configurations belonging to the states, and summing over all pairs of configurations, each one weighted with
its own statistical weight.

The self-overlap of a state is simply,
\be
q_{\a\a}=\frac1N\sum_{i=1}^N{\la\s_i\ra}_{\a}^2
\ee 
The self-overlap will be very important in what follows. It is a measure of the size of the state in 
the phase space: the larger $q_{\a\a}$, the smaller the state, i.e. the smaller the number of configurations
belonging to the state. On the other hand, a very small self-overlap indicates a very broad state. In particular,
the paramagnetic state (unbroken ergodicity) has self-overlap equal to zero. 

In the limit $T\to 0$ each states concentrate on its lowest energy configuration. In this case, the self-overlap 
of {\it each} state is $q_{\a\a}=1$, since it is just the self-overlap of a configuration. When the temperature 
$T$ grows, more configurations participate to the state and the self-overlap becomes smaller than one.

\subsection{Overlap distribution}

As we shall see, in mean-field spin-glasses there are many inequivalent pure states at low temperatures.
In this case, it is useful to introduce the probability distribution of all the possible values of the
overlaps among states. We first compute the overlap distribution by considering two physical systems
with the same disorder (also called {\it real} replicas), and averaging the value of the overlap
$q_{\s\t}$ among the two real replicas,
\be
P(q) = \frac{1}{Z^2}\int D\sigma D\tau e^{-\beta H(\sigma)} e^{-\beta H(\tau)} \ \delta(q-q_{\s\t}) \ .
\ee
Using the definitions of the previous sections, we have,
\be
P(q) = \sum_{\alpha\beta}w_\alpha w_\beta \frac{1}{Z_\alpha}\int_{\s\in\alpha}\frac{1}{Z_\beta}
\int_{\tau\in\beta}  
D\sigma D\tau e^{-\beta H(\sigma)} e^{-\beta H(\tau)} \ \delta(q-q_{\s\t}) \ , 
\ee
and using the clustering property we finally obtain,
\be
P(q) = \sum_{\alpha\beta} w_\alpha w_\beta\; \delta(q-q_{\alpha\beta}) \ .
\ee
In this formula (which can also be taken as a definition of the $P(q)$)
the sum is extended over all the possible pairs of states, including pairs of the same state, giving
its self-overlap. Once again, the simple Ising model can help us. At low temperature
we have two pure states, so we have in principle four possible overlaps,
\bea
q_{++} = \frac{1}{N} \sum_i \langle\sigma_i\rangle_+^2 = \frac{1}{N} \sum_i m_i^2 =  m^2 \\
q_{--} = \frac{1}{N} \sum_i \langle\sigma_i\rangle_-^2 = \frac{1}{N} \sum_i m_i^2 =  m^2 \\
q_{+-} = q_{-+}=\frac{1}{N} \sum_i \langle\sigma_i\rangle_+\langle\sigma_i\rangle_- = -\frac{1}{N} \sum_i m_i m_i = - m^2 \ .
\eea
Therefore the function $P(q)$ has two peaks, at$-m^2$ and $+m^2$, each with weight $1/2$. It is important to 
stress that the number of peaks of the $P(q)$ is {\it not} equal to the number of states, but to the number of
possible values taken by the overlap. If we had a very large number of states, all with the same self-overlap and
mutual overlap, we would still have a bimodal $P(q)$.

To conclude, we note that the particular structure of states of a given sample depends on the particular
realization $J$ of the quenched disorder. For this reason both the pure states weights, and the distribution
$P(q)$ depend on the disorder $J$. In particular, $P(q)$ is not a self-averaging quantity when the structure
of states is nontrivial. 
For the proof and discussion of this crucial statement see \cite{mezard84}.

\section{Statics}

We have now all the tools to start a thermodynamic study of a specific spin-glass. We will
use the replica method to compute the free energy of the system, and will discover that replicas have
(surprisingly enough) a rather deep physical meaning: they will act as probes exploring the unknown phase
space, and sending us important information on the structure of states in it.

The spin-glass model we will analyze is the $p$-{\it spin spherical model} (PSM). Among spin-glasses
it is the  one which bears more similarities with structural glasses, suggesting that some concepts which
are exactly valid for the PSM may be exported to the case of glasses. 

\subsection{The $p$-spin spherical model}

The Ising version (i.e. with $\pm 1$ spins) of the PSM was introduced in \cite{derrida}, while its spherical,
and simpler, counterpart appeared in \cite{crisanti92}. The Hamiltonian of the spherical PSM is,
\be
H=-\sum_{i_1>\ldots>i_p=1}^N J_{i_1\ldots i_p}\s_{i_1}\ldots\s_{i_p} \quad \quad p\ge 3
\ee
where the spins are now {\it real} continuous variables. In order to keep the energy finite, we have to 
put a constraint on the spins,
\be
\sum_{i=1}^N{\s_i}^2=N 
\ee
this is the {\it spherical constraint}, from which the model takes its name. With this constraint the
self-overlap of each configuration is one. The Hamiltonian is a sum of $p$-body interactions, and the
sum is extended over {\it all} groups of spins, not only the nearest-neighbor, so the model has
no spatial structure, and it is in fact a {\it mean-field} model. For such models the droplet argument given 
above does not work (each spin interacts with $N$ other spins, there are no surfaces), and thus the 
free energy barriers around metastable states may be infinite. For this reason mean-field models are 
the ideal play-ground to study metastability.

Each random coupling $J$ is a Gaussian variable, with distribution,
\be \label{gaussian_disorder}
dp(J)=\exp\left(- \frac1{2}\ J^2\ \frac{2N^{p-1}}{p!}\right)dJ
\ee
where the factors $2$ and $p!$ are a matter of convention, whereas the factor $N^{p-1}$ is 
essential in order to have the Hamiltonian of order $N$, and thus extensive energy and free 
energy,
\be
\sqrt{\ov{J^2}} \sim \frac{1}{N^{\frac{p-1}{2}}} \quad \Rightarrow \quad H \sim N
\ee

The relevance of the PSM in the context of glassy physics is due to the great role played by
metastable states in such a model. A hint of this fact comes from the ferromagnetic version
of the PSM, that is $J_{i_1\ldots i_p}=1/N^{p-1}$ for each coupling: unlike its $p=2$ counterpart,
this model has a first order transition between a high $T$ paramagnetic phase and a low $T$
ferromagnetic one (solving the ferromagnetic mean-field PSM is a trivial exercise). In 
particular, there are two relevant temperatures: a temperature $T_d$ below which a ferromagnetic
state develops, but with a free energy higher than the paramagnetic one, and a lower temperature
$T_s$, where the ferromagnetic state becomes stable and the thermodynamic transition takes place. 
From a dynamical point of view, however, the higher temperature $T_d$ is quite relevant,
since for $T<T_d$ the system may remain trapped by the ferromagnetic state, even though metastable,
if the initial magnetization is positive and large enough.

The first order transition at $T_s$ in the ferromagnetic PSM is driven by entropy, since the 
energy of the ferromagnetic states is {\it always} lower than the paramagnetic one. We can 
roughly understand this point by noting that the $p$-body interaction indeed increases very much 
the entropic contribution of the paramagnet, compared to the canonical $p=2$ case. Metastability,
entropy driven transitions, and purely dynamical transitions will be also key ingredients of the 
disordered PSM we are about to study.

\subsection{First try: the replica symmetric calculation} 

We start our static study of the PSM by performing an annealed calculation of the free energy. We know
it is wrong at low temperatures, but it will be anyway a useful warm-up exercise. In what follows
we will often write the indices for the $p=3$ case, such that $J_{i_1\dots i_p}$ becomes $J_{ijk}$.
However, to give formulas that are valid even in the general case, we will write all the factors 
containing a term $p$ for the generic $p$ case, for example we will write $N^p/p!$ rather than $N^3/6$.
Another short-cut we will use is to disregard all normalizing factors that, once taken the log and divided by
$N$, go to zero in the thermodynamic limit. Finally, we have to remember that all our integrals over
$\sigma$ are restricted to the surface of a sphere by the spherical constraint. The average partition function is given by,
\bea
\overline{Z} = \int D\sigma \; \int \prod_{i<j<k} dJ_{ijk}\; 
\exp\left[- J_{ijk}^2 \frac{N^p}{p!} + J_{ijk} \beta \sigma_i \sigma_j\sigma_k\right] = \nn \\
\int D\sigma\; \exp\left[\frac{\beta^2}{4N^{p-1}}\left(\sum_i \sigma_i^2\right)^p\right]= \nn \\
\exp\left[N\frac{\beta^2}{4}\right] \; \Omega \ ,
\eea
where $\Omega$ is the surface of the sphere. In the equations above we have used the formula,
\be
p!\sum_{i<j<k}^N = \sum_{ijk}^N
\ee
which is valid in the thermodynamic limit. The annealed free energy is therefore given by,
\be
F_a = -\beta/4 - TS_{\infty} \ ,
\ee
with the infinite temperature entropy, $S_\infty= \log(\Omega)/N$. This is, in fact, the
correct free energy at high temperatures, i.e. in the paramagnetic phase. However, it can be proved that
at lower temperatures the annealed-paramagnetic solutions has a free energy larger than the free energy
found by the quenched computation: as anticipated above the annealed approximation in general only holds
at higher temperatures, while at low temperature
the quenched computation must be performed. Note that the fact that the annealed entropy becomes
negative at low temperatures would not be by itself a sufficient reason to discard it, since
the model is continuous, and a negative entropy is thus perfectly legal.

In order to perform the quenched calculation we must compute the average of the replicated 
partition function. Since now on the indices $i,j,k,\dots$ will refer to sites, while $a,b,\dots$
will refer to replicas. We have,
\bea
\overline{Z^n} = 
\int D\sigma_i^a \prod_{ijk} \int dJ_{ijk}\;
\exp\left[- J_{ijk}^2 \frac{N^p}{p!} + J_{ijk} \beta \sum_a^n\sigma_i^a \sigma_j^a\sigma_k^a\right] = \nn \\
\int D\sigma_i^a \prod_{ijk}  \exp\left[\frac{\beta^2p!}{4N^{p-1}} \sum_{ab}^n 
\sigma_i^a\sigma_i^b \sigma_j^a\sigma_j^b \sigma_k^a\sigma_k^b\right] = \nn \\
 \int D\sigma_i^a \; \exp\left[\frac{\beta^2}{4N^{p-1}} \sum_{ab}^n \left(\sum_i^N\sigma_i^a\sigma_i^b\right)^p\right] \ .
\eea \label{tonno}
We can see here the powerful replica trick at work: we started from a set of coupled sites and uncoupled replicas,
and averaging over the disorder we decoupled the sites, but coupled the replicas (unfortunately in non-mean field 
models the replica trick is not enough to decouple the sites).
In particular, the overlap between two different replicas of the system very naturally appeared in the calculation,
\be
Q_{ab} = \frac{1}{N}\sum_i \sigma_i^a\sigma_i^b
\ee 
Note that $Q_{aa}=1$ due to the spherical constraint. 
We introduce now a factor $1$ in our calculation,
\be
1 = \int dQ_{ab}\; \delta\left(NQ_{ab} - \sum_i\sigma_i^a\sigma_i^b\right) \ ,
\ee
and finally we use an exponential representation for the $\delta$-function, to obtain,
\bea
\overline{Z^n} &=& 
\int DQ_{ab}\; D\lambda_{ab}\; D\sigma_i^a \ 
\cdot \nn \\
&\cdot& \exp\left[\frac{\beta^2N}{4}\sum_{ab}Q_{ab}^p + N \sum_{ab} \lambda_{ab} Q_{ab} - 
\sum_i\sum_{ab} \sigma_i^a\lambda_{ab}\sigma_i^b \right] = \nn \\
&=&\int DQ_{ab}\; D\lambda_{ab}\; \exp\left[-N\; S(Q,\lambda)\right] 
\label{into}
\eea
with,
\be
S(Q,\lambda) = - \frac{\beta^2}{4}\sum_{ab}Q_{ab}^p - \sum_{ab} \lambda_{ab} Q_{ab} +\frac{1}{2}\log\det(2\lambda_{ab})
\ee
In (\ref{into}) the integration over $Q_{ab}$ is performed over all the matrices with $a\neq b$, while the integration
over $\lambda_{ab}$ includes also $a=b$ to enforce the spherical constraint. The sums in the exponentials are
over all the indices, including $a=b$.

The great advantage of this form of the integral is that we can use the saddle point (or Laplace, or steepest-descent) method 
\cite{bender-orszag}, to solve it in the limit $N\to\infty$. 
This simplification is the {\it big} effect of mean-field, and it is the result of 
the decoupling of the sites operated by the use of the
replica trick. The price we had to pay is that we coupled replicas, and this looks somewhat weird at this
stage of the computation. 

The saddle-point method states that in the limit $N\to\infty$ the integral (\ref{into}) is concentrated in the 
minimum of the integrand. However, we have to be careful here, for a twofold reason. First, the free energy is 
in principle given by,
\be
-\beta F = \lim_{N\to\infty}\lim_{n\to 0} \frac{1}{nN} \log \int DQ_{ab}\; D\lambda_{ab}\; \exp\left[-N\; S(Q,\lambda)\right] 
\ee
and thus we should {\it first} take the limit $n\to 0$, and then $N\to\infty$. Unfortunately, we are unable to do this:
$S$ is not an explicit function of $n$, and moreover we need to send $N\to\infty$ first to solve the integral.
As a conclusion, we need to exchange the order of the two limits, solve the integral, find a parametrization of the
matrix $Q_{ab}$, and finally take the $n\to 0$ limit at the end. Of course, this is mathematically risky, to say the least.

The second point we have to pay attention to, is what do we actually mean by ``minimum'' of $S$. The problem here is that
the number of independent elements of $Q_{ab}$ is $n(n-1)/2$, which becomes negative is the limit $n\to 0$. It is hard
to say what is a minimum of a function with a negative number of variables ! There is however a criterion we can use
to select the correct saddle point: the corrections to the saddle point result are given by the Gaussian integration
around the saddle point itself. This integration gives as a result the square root of the determinant of the second
derivative matrix of $S$, and thus, in order to have a sensible result, we must have all the eigenvalues of this
matrix positive. Summarizing, we have to select saddle points with a positive-defined second derivative of $S$
\cite{dealmeida78}.

At this point we can proceed with the saddle point calculation. We first minimize (maximize ?) $S$ with respect to 
$\lambda_{ab}$. By using the general formula,
\be
\frac{\partial}{\partial M_{ab}}\; \log \det M_{ab} = (M^{-1})_{ab}
\ee
we get,
\be
2\lambda_{ab} = (Q^{-1})_{ab}
\ee
and thus,
\be
F = \lim_{n\to 0} -\frac{1}{2\beta n}\left[ \frac{\beta^2}{2}\sum_{ab} Q_{ab}^p + \log\det Q_{ab}\right]
\label{fifa}
\ee
where $Q_{ab}$ satisfies the saddle point equation,
\be
0 = \frac{\partial F}{\partial Q_{ab}} = \frac{\beta^2 p}{2} Q_{ab}^{p-1} + (Q^{-1})_{ab} \ .
\label{sadu}
\ee
Note that $Q_{aa}=1$ due to the spherical constraint.

What we have obtained is a free energy $F$, function of an order parameter, $Q_{ab}$, which is 
definitely weirder than the simple magnetization $m=\langle \sigma\rangle$ we would have in the ferromagnetic Ising model. This 
order parameter is the overlap between configuration belonging to different replicas, and its physical meaning will 
be clearer later on. For now, we limit ourselves to find a solution of the saddle point equation. To do this we have first 
to find a parametrization of the matrix $Q_{ab}$, and to write (\ref{sadu}) as a function of the elements of $Q_{ab}$ and of
its dimension $n$.

Given that all replicas are equivalent (they just come from a formal trick !), it seems wise to assume 
a {\it replica symmetric} form for the matrix $Q_{ab}$. This is what Sherrington and Kirkpatrick did in their first
mean-field spin glass model \cite{sherrington75}, that is,
\be
Q_{ab} = q_0 + (1-q_0)\delta_{ab}  \ .
\ee
This means that all the elements of $Q_{ab}$ are equal to $q_0$, but on the diagonal, where they are $1$. The value of
$q_0$ must be found from the saddle point equations. We have,
\be
(Q^{-1})_{ab} = \frac{1}{1-q_0}\delta_{ab} - \frac{q_0}{(1-q_0)[1+(n-1)q_0]} 
\ee
and thus (\ref{sadu}) becomes, in the limit $n\to 0$,
\be
\frac{\beta^2 p}{2} q_0^{p-1} - \frac{q_0}{(1-q_0)^2} = 0 
\ee
We first see that $q_0=0$ is always solution of this equation. This is the paramagnetic solution, and from 
(\ref{fifa}) we get $F = -\beta/4$, which is the same as the annealed result (except for the phase space 
volume $S_{\infty}$ we did not include here). Thus, the annealed calculation gives the same result as the
quenched calculation when the overlap matrix $Q_{ab}$ is the identity. This is obvious, because when 
$Q_{ab}=\delta_{ab}$ replicating or not the system is exactly the same.

However, we also have a non-paramagnetic solution $q_0\neq 0$. Recasting the equation in the following form,
\be
q_0^{p-2}(1-q_0)^2 = \frac{2}{p} T^2
\label{pina}
\ee
we clearly see that at high temperatures there is no nontrivial solution, while by decreasing $T$ we arrive at a critical
value $T^\star$ below which a pair of nonzero solutions forms. Of these two solution the only acceptable one is the 
larger one, which increases with decreasing $T$ (the self overlap must increase if the number of
configurations belonging to a state decreases, and this is exactly what we expect when we decrease the 
temperature). 
Therefore we seem to have a transition at $T^\star$, and in particular a {\it discontinuous} transition, since the 
value of $q_0$ at the transition is different from zero, i.e. there is a jump of the order parameter at the transition.
Moreover, the free energy associated to this new solution is lower than the paramagnetic one, therefore it would
seem we have found the new non-paramagnetic state at low temperatures.

All this seems very interesting, but there is a problem: the nontrivial solution we have found is {\it unstable}
\cite{dealmeida78,crisanti92}.
As we have said above, when we select a saddle point, we have to be sure that all the eigenvalues of the second 
derivative of $F$ around the saddle point are positive. Unfortunately, this is not the case for this solution: both
roots of equation (\ref{pina}) have one negative eigenvalue below $T^\star$.

What can we do ? Remember that we did not search the whole space of $Q_{ab}$ to find a solution, but rather assumed
a certain parametrization, which looked more or less sensible, and plugged it into the saddle point equation.
The fact that the replica symmetric ansatz gave us a nontrivial solution at low $T$, but which is unstable,
clearly means that the low temperature phase of the model must be describe by a {\it replica symmetry breaking}
form of the order parameter $Q_{ab}$. Before looking for this new solution, it is finally the moment to try and understand 
what is the physical meaning of the weird order parameter $Q_{ab}$.

\subsection{The key connection between replicas and physics}

Let us consider the following quantity,
\be
q^{(1)}= \frac{1}{N}\sum_i \overline{\langle\sigma_i\rangle^2}
\ee
which, as we have seen in the previous chapter,  is a quite natural definition of an order parameter, 
since it is just a generalization  of the average  magnetization $m$.
By using the technology developed in the first chapter, we can rewrite $q^{(1)}$ in the following way,
\bea
q^{(1)}= 
\frac{1}{N}\sum_i\sum_{\alpha\beta} \overline{w_\alpha w_\beta\  \langle\sigma_i\rangle_\alpha \langle\sigma_i\rangle_\beta} =
\sum_{\alpha\beta} \overline{w_\alpha w_\beta \ q_{\alpha\beta }} =\nn \\
\int dq \sum_{\alpha\beta} \overline{w_\alpha w_\beta \ \delta(q-q_{\alpha\beta})}\ q = 
\int dq \; \overline{P(q)}\; q
\eea
Therefore $q^{(1)}$ is the first moment of the overlap distribution, averaged over the disorder. By using the clustering
property, we can easily find a generalization of this formula \cite{me-pa-vi},
\be
q^{(k)} = \frac{1}{N^k}\sum_{i_1\dots i_k} \overline{\langle\sigma_{i_1}\dots \sigma_{i_k}\rangle^2} = 
\int dq\;  \overline{P(q)}
\; q^k \label{picco}
\ee
The important fact is that we can compute these quantities also using the replica trick. In particular,
\be
q^{(1)}= \frac{1}{N}\sum_i \overline{\langle\sigma_i\rangle^2} = 
\lim_{n\to 0}\overline{
\int D\sigma^a_i \ \frac{1}{N} \sum_i \sigma_i^1\cdot\sigma_i^2 \ e^{-\beta\sum_a H(\sigma^a)}
}
\ee
If we now go on with the calculation along the lines of the previous paragraphs, introducing the 
overlap matrix $Q_{ab}$, we get,
\be
q^{(1)}= \int DQ_{ab} \;\; e^{-NS(Q_{ab})}\, Q_{12} = Q_{12}^{\rm{SP}}
\label{pluto}
\ee
where $Q_{ab}^{\rm{SP}}$ is the saddle point value of the the overlap matrix (since now on we will drop the 
suffix SP), and where we have exploited the fact that $S$ is of order $n$, and therefore does not contribute when 
$n\to 0$. Of course, there is something wrong about this formula: replicas 1 and 2 cannot be different from the
others! If we decided to call them 4 and 7, we would get a different result when $Q_{ab}$ is not replica symmetric:
this is nonsense ! What is going on here ? To understand this point we note that if the saddle point 
overlap matrix is not symmetric, then there must be other saddle point solutions with the same free energy, 
but corresponding to matrices obtained from $Q_{ab}$ by a permutation of lines and columns \cite{me-pa-vi}. 
This is a general result: when a saddle point
breaks a symmetry corresponding to a given transformation, all the points obtained by applying the transformation
to that particular saddle point, are equally valid. This means that we must average over all these
saddle points, and this is equivalent to symmetrize equation (\ref{pluto}) \cite{parisi83, dedominicis83}, 
obtaining,
\be
q^{(1)} = \lim_{n\to 0} \frac{2}{n(n-1)} \sum_{a>b} Q_{ab} 
\label{sunto}
\ee
This result is already telling us that there is a connection between the physical order parameter
$q^{(1)}$, and the matrix of the overlap among replicas $Q_{ab}$. To go further, we can generalize (\ref{sunto}), 
to get,
\be
q^{(k)} = \lim_{n\to 0} \frac{2}{n(n-1)} \sum_{a>b} Q_{ab}^k 
\label{sunto2}
\ee
A comparison with equation (\ref{picco}), gives for a generic function $f(q)$ the  relation,
\be
\int dq\; f(q)\;\overline{P(q)} = \lim_{n\to 0} \frac{2}{n(n-1)} \sum_{a>b}f(Q_{ab})
\ee
and in particular choosing $f(q)=\delta(q-q')$, we finally find the crucial equation connecting physics to
replicas,
\be
\overline{P(q)} = \lim_{n\to 0} \frac{2}{n(n-1)} \sum_{a>b} \delta(q-Q_{ab}) \ .
\label{funda}
\ee
This equation is telling us that the average probability that two pure states of the system have overlap
$q$ is equal to the  fraction of elements of the overlap matrix $Q_{ab}$ equal to $q$. In 
other words, {\it the elements of the overlap matrix (in the saddle point) are the physical values 
of the overlap  among pure states, and the number of elements of $Q_{ab}$ equal to $q$ is related to the 
probability of $q$}.

This is a key connection, and we understand now that $Q_{ab}$ has an enormous
physical meaning. As a first application, let us analyze the meaning of the replica symmetric
ansatz, $Q_{ab}=q_0$ for each $a\neq b$. From (\ref{funda}) we see that this structure of
the overlap matrix implies that the average overlap distribution is given by,
\be
\overline{P(q)} = \delta(q-q_0) \ ,
\ee
that is there is one single possible value of the overlap among states. As we have seen, the
overlap distribution should also include the self-overlap of the states, and therefore
this value $q_0$ must be the
self-overlap of the unique state in the system. The conclusion is that a replica symmetric form of the
overlap matrix in the free energy calculation, can only be valid if there is one single 
equilibrium state. This state will typically be the paramagnetic state, and its self-overlap will be
$q_0$. On the other hand, if at low temperatures there is ergodicity breaking, with the emergence  of 
many inequivalent pure states, then the correct form of $Q_{ab}$ {\it cannot} be replica symmetric.

Now that we know what is the meaning of the overlap matrix, there is a slight chance 
to understand how to find a replica symmetry breaking form of it.

\subsection{Replica symmetry breaking}

Thanks to equation (\ref{funda}) the relations of overlap among states translate into relations of
overlaps among replicas. Therefore, in order to give an ansatz on the form of $Q_{ab}$ we have to 
guess what may be the structure of states in the low temperature phase of a spin-glass model. 
{\it Vast programme !} -  as someone once said \cite{degaulle}.

Our starting point is a fact we already know: if there is ergodicity breaking, that is if there are
many states, configurations in the phase space are organized  into states. In other words, we can
think of states as blobs of configurations in the phase space, with each configuration belonging to 
just one blob. The self overlap of a state is just the average overlap of the configurations belonging 
to it, i.e. it is a measure of the largeness of the blob. On the other hand, the overlap between different
states is basically the overlap between configurations belonging to them. 

Given this, the simplest
possible spectrum of overlaps we can have when there are many states is the following: $q=1$, if we
consider twice the same configuration, $q=q_1<1$ if we consider different configurations belonging to 
the same state, $q=q_0<q_1$ if we consider configurations belonging to different states. In this way we
are assuming that all states have the same self-overlap $q_1$, and mutual overlap $q_0$. Moreover, 
a physical requirement is that $q_1 > q_0$, since configurations belonging to the same state must be
closer than those in different states.

What is the corresponding structure of $Q_{ab}$ ? What is clear by now is that replicas act as probing
configurations of the structure of the states, so we must reproduce for replicas the same clustering
procedure we have seen for the configurations: replicas may belong to the same group, having overlap
$Q_{ab}=q_1$, or to different groups, with overlap $Q_{ab}=q_0$. Finally, when we select twice the same 
replica we obtain $Q_{aa}=1$. To this structure corresponds the matrix \cite{parisi79},
\be
Q_{ab}=
\left[
\begin{array}{ccc}
\begin{array}{ccc}
1 & q_1 & q_1 \\ q_1 & 1 & q_1 \\ q_1 & q_1 & 1
\end{array}
& q_0 & \cdots \\
q_0 &
\begin{array}{ccc}
1 & q_1 & q_1 \\ q_1 & 1 & q_1 \\ q_1 & q_1 & 1
\end{array}
& \\
\vdots & & \ddots
\end{array}
\right]
\ee
where we have assumed, to make an example, that the number $m$ of replicas in each group is $m=3$. As we have seen, the parameter
$m$ is connected to the probability of having a given value of the overlap, therefore it will become a variational
parameter in the saddle point equations, as $q_1$ and $q_0$. This structure of $Q_{ab}$ reflects what we have said 
above. Note that any permutation of lines or column (replica permutation) would also correspond to the same structure,
but it would simply be much harder to visualize. This matrix has the important property that $\sum_a Q_{ab}$ does 
not depend on $b$, which is an essential requirement, since replicas must be all equivalent \cite{parisi80b}.

It is clear that the clustering process we have described can be iterated \cite{parisi80}: 
states can be grouped into clusters, which 
can be grouped into super-clusters, and so on. The structure of states one obtains in this way is called {\it ultrametric},
and unfortunately we do not have time to describe it here \cite{mezard84}. The important point is that 
 for the PSM the simple structure described above is sufficient \cite{crisanti92}. 
This kind of replica symmetry breaking (RSB) is called {\it one step} RSB, or 1RSB.

Let us compute the overlap distribution associated to the 1RSB structure of $Q_{ab}$. From (\ref{funda}) we get,
\be
\overline{P(q)} = \frac{m-1}{n-1}\, \delta(q-q_1) + \frac{n-m}{n-1} \,\delta(q-q_0)
\label{gaio}
\ee
with,
\be 
1 \leq m \leq n
\label{bigu}
\ee
We clearly see that while the elements $q_1, q_0$ fix the position of the peaks, 
the parameter $m$ fixes their height, and therefore the probability of the overlaps. 
We have now to take the limit $n\to 0$. Here lies possibly the weirdest twist
of the replica method. Relation (\ref{bigu}) seems to resist strenuously to our will
to send $n$ to zero. However, in this limit it is clear that also $m$ must be promoted
to be a real number, rather than an integer. To see how to do this we can appeal to
physics (for once !), and accept the fact that the {\it probability} (\ref{gaio}) must be positive
even in the limit $n\to 0$,
\be
\overline{P(q)} = (1-m) \, \delta(q-q_1) + m \, \delta(q-q_0)
\label{narciu}
\ee
For this to be positive we must have $m<1$ and $m>0$. Therefore, the correct limit of
(\ref{bigu}) for $n\to 0$, is,
\be
0 \leq m \leq 1 \ .
\ee
Summarizing, with the 1-step replica symmetry breaking ansatz we have parametrized the overlap 
matrix $Q_{ab}$ by means of two values of the overlap,
\be
0 \leq q_0 \leq q_1 \leq 1
\ee
and one value of the probability parameter $m$. We have now to fix them via the saddle point equation.

\subsection{The 1RSB solution and the static transition}

The first thing to do is to compute the free energy as a function of $q_1, q_0, m$. We had,
\be
F=\lim_{n \rightarrow 0}-\frac1{2\b n}\left[\frac{\b^2}2\sum_{a,b}Q^p_{ab}+\log\det \, Q\right]
\ee
The first piece is easy to compute and gives in the limit $n\to 0$,
\be
\frac{1}{n}\sum_{ab} Q_{ab}^p = \sum_a Q_{ab}^p = 1 + (m-1) q_1^p - m q_0^p
\ee
The second piece is a bit harder: the 1RSB matrix $Q_{ab}$ has three different eigenvalues and
degeneracies (the student should be able to work them out),
\bea
\lambda_1 = 1-q \quad\quad d_1 = n-n/m \\
\lambda_2 = m (q_1-q_0)+ (1-q_1)  \quad\quad d_2 = n/m -1  \\
\lambda_3 = n q_0 + m (q_1-q_0)+ (1-q) \quad\quad d_3 = 1
\eea
From this, taking {\it carefully} the limit $n\to 0$, we finally obtain,
\bea
\label{f1rsb}
-2\b F_{1RSB}&=&\frac{\b^2}2[1+(m-1)q^p_1-mq^p_0]+\frac{m-1}m\log(1-q_1)+ \nn \\ 
&+&\frac1m\log[m(q_1-q_0)+(1-q_1)]+\frac{q_0}
{m(q_1-q_0)+(1-q_1)}
\eea
to be compared with the replica symmetric (RS) free energy,
\be
-2\b F_{RS} = \frac{\b^2}2[1 - q^p_0]+ \log(1-q_0) + \frac{q_0}{1-q_0}
\ee
It is interesting to note that the RS form is obtained either from $q_1\to q_0$, or $m\to 1$. In the first limit,
many states merge to form a single paramagnetic state. This is exactly what happens in the Ising model when $T\to T_c^-$,
and it is a consequence of the second order nature of the transition in that model. The $m\to 1$ limit has a different,
less trivial, interpretation, as we shall see in a minute.

We have now to study the saddle point equations with respect to $q_1, q_0, m$.
First, the equation $\partial_{q_0} F=0$ gives as a solution $q_0=0$. This solution is correct in absence of
external magnetic field: $q_0$ is the mutual overlap among different states, and it is natural to think that
without external field  the distribution of states in the phase space is symmetric, and thus all states must be
orthogonal to each other. 

The two remaining equations  $\partial_{q_1} F=0$ and $\partial_{m} F=0$ are,
\bea
(1-m)\left( \frac{\beta^2}{2} p q_1^{p-1} - \frac{q_1}{(1-q_1)[(m-1)q_1+1]}\right) &=& 0 \nn \\
 \frac{\beta^2}{2} q_1^{p} + \frac{1}{m^2}\log\left(\frac{1-q_1}{1-(1-m)q_1}\right) + \frac{q_1}{m[1-(1-m)q_1]} &=& 0
\eea
These equations can be easily studied on a computer, but most of the physics can be worked out also 
graphically. At high $T$ the only solution is $q_1=0$ and $m$ undetermined: this is the paramagnetic 
solution, which is equal to the RS one. We want to know whether there is a nontrivial spin-glass solution
with $q_1\neq 0$. The first equation is solved by $m=1$. So let us plug $m=1$ into the
second equation,
\be
\frac{\beta^2}{2} q_1^{p} + \log\left(1-q_1 \right) + q_1 \equiv g(q_1) = 0
\ee
The graphical study of this equation for $0\leq q_1\leq 1$ is trivial. The limits are $g(0)=0$ and $g(1)=-\infty$.
At high $T$ the function is monotonous and only the $q_1= 0$ solution exists.
However, by lowering the temperature, $g(q_1)$ develops a maximum, whose height diverges for decreasing $T$.
Therefore, it must exist a temperature $T_s$, where this maximum touches the axis at $q_1 \equiv q_s \neq 0$. 
Therefore, at $T=T_s$ a new spin-glass solution appears, with 
$q_1=q_s$ and $m=1$. When $T< T_s$ we have to move $m$ from $1$, and one can see
that the solution simply shifts, $q_1>q_s$, $m<1$ \cite{crisanti92}.
The important point is that, unlike the RS nontrivial solution, this solution is {\it stable}. 
Moreover, its free energy its lower than the paramagnetic one. 
The temperature $T_s$ where this nontrivial 1RSB solution appears is called {\it static transition}
temperature.  It has been proved that 
the 1RSB solution is {\it exact} in the PSM \cite{crisanti92}. This means that if we take
higher order RSB ansatz for $Q_{ab}$, from the saddle point equations we find that all the 
extra parameters we introduce have in fact a trivial value, and that the 1RSB solution is recovered.

As we have seen, at $T_s$ the value of the self-overlap is nonzero, $q_1=q_s$, 
while $m=1$. This fact has an  interesting physical interpretation. For $T>T_s$ the overlap distribution is 
trivial, $P(q)=\delta(q)$. By decreasing the temperature, finally a second peak appears at $T_s$, for $q=q_s$.
This value is nonzero, meaning that at the transition the states are already well formed, that is tight.
However, at $T_s$ we have $m=1$, meaning that the height of this nontrivial peak is in fact zero:
the {\it probability} of these new states is zero at the transition, and it grows below $T_s$ as $m$ becomes smaller 
than one. In other words, it seems that when the states appear, they are already well formed, but have a zero
thermodynamic weight. 

This fact has a possible interpretation in terms of metastable states: the calculation
we have just performed is a {\it thermodynamic} one, and therefore by its very nature it is unable to 
capture the contribution of metastable states. If in this systems there were some metastable states even 
above $T_s$, and some of them became stable only  below $T_s$, the behaviour of $P(q)$ would be exactly 
the one described above. The states are already present in the phase space, even above $T_s$,
with a well defined nonzero self-overlap, but their thermodynamic weight is zero, as long as the temperature
does not drop below the static transition. At that point the free energy of these states becomes smaller
than the paramagnetic one, therefore their weight is nonzero, and the $P(q)$ develops a secondary peak.
For now, this is just a well motivated hypothesis. We will see in the following chapters that
it is in fact verified.

Summarizing, in the PSM we find a static transition between a high temperature paramagnetic phase, 
and a low temperature spin-glass phase below $T_s$. In this phase many pure states dominate the
partition function. The order parameter of this unusual transition is the overlap matrix $Q_{ab}$, and
more precisely, within the 1RSB scheme, the self-overlap $q_1$ and the probability parameter $m$. In the 
paramagnetic phase the overlap matrix has a  replica symmetric form, with $q_1=0$ and $m$ undetermined,
while in the spin-glass phase there is replica symmetry breaking, with nontrivial values of $q_1$ and $m$.
The nature of this transition is discontinuous if we consider the parameter $q_1$, but continuous if we
consider the whole probability distribution  $P(q)$.

\section{Equilibrium dynamics}

As we have already pointed out, the dynamical behaviour of a system will be very different from its thermodynamic
behaviour, if metastable states are present. This is particularly true in mean-field, where metastable states may have 
infinite lifetime.  The results from the previous chapter seem to suggest that something nontrivial is going on in the
PSM even for $T>T_s$. It is therefore important  that we perform an independent dynamical study of the model,
and see whether our guess about the presence of metastable states was right. A nice introduction to the main concepts 
of dynamics can be found in \cite{zwanzig}.

\subsection{The generating functional formalism}

Before focusing on the PSM, we give a brief summary on how to study the dynamics of a degree of freedom $x(t)$,
described by an Hamiltonian $H(x)$,  which contains some quenched disorder.
The starting point of our dynamical study is the Langevin equation \cite{gardiner},
\be
\frac{d x}{dt}=-\frac{\partial H}{\partial x}+\eta(t)
\ee
where $\eta(t)$ is a Gaussian noise, playing the role of the thermal agitation, with
\be
\la \eta(t)\ra=0 
\ee
\be
\la \eta(t)\eta(t')\ra=2T\;\d(t-t') 
\ee
The factor $2T$ is crucial, since it relates the strength of the noise to the friction coefficient in the 
Langevin equation \cite{zwanzig}.
In the rest of this chapter we will have to integrate repeatedly  over the degrees of freedom $x(t)$, and over the 
disorder $\eta(t)$, which are both functions of time. Thus, most of the integrals will be {\it functional integrals}. 
With the notation $Dx$ we actually mean a measure over all the paths, i.e. $D[x(t)]$ \cite{zinn}.
The probability $P(\eta)$ of the noise can be written as,
\be
P(\eta)\sim\exp\left[-\frac12\int dt dt' \,\eta(t)D^{-1}(t-t')\eta(t')\right]
\ee
with
\be
D(t-t') = 2T \; \delta(t-t')
\ee
Every solution $x(t)$ of the Langevin equation depends on the particular realization of the thermal noise $\eta(t)$, 
and we indicate it as $x_{\eta}(t)$. 
From the probability distribution on $\eta$ we can therefore obtain a distribution on $x$. To work this out,
let us compute the average over the noise of a generic observable $A$, function of the degree of freedom $x(t)$,
\bea
\la A(x)\ra &=&\int D\eta\, P(\eta)\, A(x_\eta)=
\int D\eta\, P(\eta)\int dx \, \d(x-x_\eta)\, A(x)= \nn \\
&=& \int dx \left[ \int D\eta\, P(\eta)\, \d(\p_tx+\p_xH-\eta) \right] A(x)=\nn \\
&=&\int dx\,  P(x) A(x) 
\eea
where we have defined the probability of $x$ as,
\be
P(x) =  \int D\eta\, P(\eta)\, \d(\p_tx+\p_xH-\eta)
\ee
Note that in principle we should introduce the Jacobian of the equation in the formula above. However, it can
be proved that if we discretize the Langevin equation according to the Ito prescription, this Jacobian is in fact equal
to $1$, and can therefore be neglected \cite{gardiner, vankampen}.

When the Hamiltonian contains quenched disorder $J$ (as in the case of the PSM) we must find a way to average over $J$.
Clearly, if we simply average the Langevin equation we get a disappointing $0=0$. In fact, the correct quantity 
which has to be averaged over the disorder is the distribution $P(x)$ above, such that to compute the average
over $J$ of an observable $A$ we can simply use the formula,
\be
\overline{\langle A(x)\rangle} = \int Dx\ \overline{P(x)} \ A(x)
\ee
Rather than precisely compute $\overline{P(x)}$, we average the
integral of $P(x)$, in order to deal with a scalar quantity. This is the generating functional method \cite{msr,janssen}, 
which has been first applied and studied in the field of spin-glasses in \cite{dedo1, dedo2, peliti}.
The starting point of the method is an apparently redundant 
way to represent the number $1$,
\bea
1&\equiv& Z = \int Dx \, P(x) 
=\int Dx D\eta \, P(\eta)\, \d(\p_tx+\p_xH-\eta) = \nn \\
&=& \int Dx D\hat{x} D\eta \, \exp\left[-\frac12\int dt dt' \, \eta(t) D^{-1}(t,t')\eta(t')+ \right. \nn \\
 &+& i \int dt \, \hat{x}(t)
(\p_t x+\p_x H)-i \int dt \, \hat{x}(t)\eta(t)\left.\right] \nn \\
&=& \int Dx D\hat{x} \, \exp\left[-\frac12\hat{x}D\hat{x}+i\hat{x}(\p_tx+\p_xH)\right] \nn \\
&\equiv& \int Dx D\hat{x} \, \exp\left[ S(x,\hat x)\right]
\label{potus}
\eea
with $S=-\frac12\hat{x}D\hat{x}+i\hat{x}(\p_tx+\p_xH)$.
We used the integral representation of the delta function and the fact that the functional integral is Gaussian. In
our notation we do not indicate explicitly the time contractions: 
$\hat{x}D\hat{x}=\int dt \; dt'\, x(t)D(t,t')x(t')$, and the same holds for $i\hat{x}(\p_tx+\p_xH)$. 

The quantity $Z$ is
the generating functional. The fact that it is just equal to $1$, must not deceive the student. In fact, we can calculate
all the interesting dynamical quantities with this functional. Let us see how. When we introduce a time dependent 
magnetic field in the system, we have an extra term,
\be
\int dt \ x(t) h(t)
\ee
in the Hamiltonian. Thus we have an extra term $h(t)$ in the original Langevin equation, which translate
into a term,
\be
\int dt \ \hat x(t) h(t)
\ee
in the action $S$ above. Therefore, when we derive the average of any quantity with respect to $h(t)$ we 
pull down a factor $\hat x(t)$ from the exponential, and in particular,
\be
\frac{\p}{\p h(t)} \la x(t') \ra = \la \hat x(t) x(t')\ra \equiv R(t,t')
\ee
which is the dynamical response function of the system, i.e. the dynamical equivalent of the susceptibility
in thermodynamics. On the other hand, we can couple a field $\hat h(t)$
to $x(t)$ in the generating functional, and get,
\be
\frac{\p}{\p \hat h(t)} \la x(t') \ra = \la x(t) x(t')\ra \equiv C(t,t')
\ee
that is the time dependent correlation function. Summarizing, once the conjugate fields $h(t)$ and
$\hat h(t)$ are introduced, we have,
\bea
R(t,t') &=& \frac{\p Z}{\p \hat h(t') \p h(t)} \nn \\
C(t,t') &=& \frac{\p Z}{\p \hat h(t') \p \hat h(t)}
\eea

Let us now consider a system with quenched disorder in the Hamiltonian. 
We first define,
\be
{\cal L}(x) \equiv \p_t x+\frac{\p H}{\p x}
\ee
and then split the Hamiltonian into a part without disorder $H_0$, and a part with disorder $H_J$,
such that the Langevin equation becomes,
\be
{\cal L}(x)={\cal L}_0(x)+{\cal L}_J(x)= \eta(t)
\ee
with
\be
{\cal L}_0 = \p_t x+\frac{\p H_0}{\p x} \quad \quad \quad {\cal L}_J = \frac{\p H_J}{\p x}
\ee
and
\be
\la \eta\eta\ra=2T\delta(t-t') \equiv D_0(t-t') \ .
\ee
The generating functional becomes,
\be
Z=\int Dx D\hat{x}\exp\left\{-\frac12\hat{x}D_0\hat{x}+i\hat x \; \left[ {\cal L}_0(x)+ {\cal L}_J(x)\right] \right\}
\label{zuk}
\ee
It should be clear by now that $Z$ has, in the dynamical approach,  the same role as the partition function in 
thermodynamic. This may suggest that, when averaging over $J$, we should consider $\log\; Z$ rather than $Z$, in 
order to reproduce the quenched case. However, this is not the case. In fact, the crucial  point
is that $Z=1$, and thus it can be safely averaged over $J$ ! Therefore in the dynamic approach we do not need
replicas \cite{dedo2}. This does not mean that the calculation will be simpler. Actually, we will see that {\it time} plays the same role as {\it replicas}: by averaging over $J$ we will decouple the sites, but couple different times.

We fact that replicas are not needed in the dynamical case can be understood also in a more direct way.
As we have already said, the correct quantity to average over the disorder $J$ is the probability distribution
$P_J(x)$ of the degree of freedom $x$. In the static case we have,
\be
P^{(s)}_J(x) = \frac{e^{-\b H_J(x)}}{\int D\hat x\; e^{-\b H_J(\hat x)}}
\ee
In order to be averaged over $J$ this expression must be rewritten in terms of replicas,
\be
P^{(s)}_J(x) = \lim_{n\to 0} \int D\hat x_{a\neq 1} \ e^{-\b \sum_a^n\;  H_J(\hat x_a)} \ ,
\ee
where $\hat x_1=x$. 
On the other hand, from (\ref{potus}) we see that the distribution in the dynamic case is given by,
\be
P^{(d)}_J(x)= \int  D\hat x(t) \ e^{\int dt \; S_J[x(t),\hat x(t)]}
\ee
where we have reinstated the dependence on the time $t$ to emphasize the formal similarity between
the replicated static case and the unreplicated dynamic one. It is clear that in the latter case we do not 
need replicas to average over $J$. However, in the dynamic case the variable $t$ plays a role analogous to
the replica index $a$. 

Expression (\ref{zuk}) is interesting for two reasons: first, the coefficients of $\hat x^2$ and $\hat x$
are respectively the correlator of the noise, and the noise-independent part of the original Langevin
equation; second, the disorder $J$ is only contained in ${\cal L}_J$ at the exponent, and it can be easily 
integrated out. This average will {\it renormalize} the coefficients of $\hat x^2$ and $\hat x$,
giving rise to a new effective Langevin equation \cite{kirkpatrick87}. More specifically, the average over $J$ gives,
\be
\overline{Z}=\int Dx D\hat{x}\exp\left[-\frac12\hat{x}D_0\hat{x}+i\hat{x}{\cal L}_0(x)\right]\overline{\exp[
i\hat{x}{\cal L}_J(x)]}
\ee
We define the quantity $\Delta(x,\hat{x})$ as  
\be
\exp[\Delta(x,\hat{x})]\equiv\overline{\exp[i\hat{x}{\cal L}_J(x)]}
\ee
Once the average is done, it is possible in general to isolate various pieces in $\Delta$, and in particular,
\be
\Delta(x,\hat{x})= -\frac12 \hat{x}D_1(x,\hat{x})\hat x+i\hat{x}{\cal L}_1(x,\hat{x})+\ldots
\ee
where ${\cal L}_1$ renormalizes the disorder-independent part of the Langevin equation ${\cal L}_0$, and $D_1$ renormalizes 
the noise correlator. In the end we have the effective Langevin equation, 
\be
{\cal L}_0(x)+{\cal L}_1(x,\hat{x})=\xi \qquad \mbox{with} \quad \la\xi\xi\ra=D_0+D_1(x,\hat{x})
\ee
In this equation the disorder is no 
longer present, but we had to pay a price: the original equation gets some nontrivial corrections. The most evident 
difference is that the variable $\xi$, 
the new effective noise, is no longer delta-correlated in time. 
In other words the integration over $J$ has introduced a sort of memory in the dynamics of the system. This phenomenon is
common in statistical physics: whenever starting from a Markovian stochastic process we integrate over some degrees of 
freedom (the disorder, the fast variables, the momenta, etc.), we end up with a new effective equation which is no longer 
Markovian, and where modes which were previously uncoupled, are now coupled (a simple example of this phenomenon
can be found in \cite{zwanzig}).

\subsection{Dynamics of $p$-spin spherical model}

In the following section we will apply the technique described above to the PSM \cite{kirkpatrick87,crisanti93}.
The formalism is trivially generalized to the case of a vectorial degree of freedom $\s_k$.
In the Langevin equation we must add a Lagrange multiplier $\mu(t)$ in order to impose the spherical constraint:
\be
\p_t\s_i(t)=-\frac{\p H}{\p \s_i}-\mu(t)\s_i(t)+\eta_i(t)\quad \mbox{with} \quad
\la \eta(t)\eta(t')\ra=2T\d(t-t')
\ee
The derivative of the Hamiltonian with respect to $\s_i$ gives,
\be
\frac{\p H}{\p \s_i}=-\frac{p}{p!}\sum_{kl}J_{ikl}\; \s_k \s_l 
\ee
The generating functional is given by equation (\ref{zuk}), with $x\to \s_k$, $\hat x\to \hat \s_k$, and,
\bea
i\hat\s \cdot {\cal L}_0 &=& \sum_k \int dt \ i\hat{\s_k}(t) \; [ \p_t\s_k(t)+\mu(t)\s_k(t) ]    
\nn \\
i \hat \s\cdot {\cal L}_J&=& -\frac{ip}{p!}\int dt \sum_{ikl} J_{ikl} \; \hat{\s_i}(t)\s_k(t)\s_l(t)
\label{summa}
\eea
If we compare this last expression with the static formulas (see equation \ref{tonno}), we can see that in this case the time has the same function as the
replica index,
\be
\int dt \, \sum_{ikl} J_{ikl}\   \hat{\s}_i(t)\s_k(t)\s_l(t)
 \longleftrightarrow 
 \sum_a \sum_{ikl} J_{ikl} \ \s^a_i \s^a_k \s^a_l
\ee
The following step is to average over the disorder, that is to compute $\overline{\exp(i\hat\s\cdot {\cal L}_J)}$.
In the statics this operation gives a coupling among replicas, in this case we will have a coupling among times. A
technical remark: before averaging, we need to symmetrize the term $\hat\s\s\s$ in (\ref{summa}), since the couplings are
completely symmetric. We find, 
\bea
&& \overline{\exp(i\hat\s\cdot {\cal L}_J)} = \nn \\
&& = \int \prod_{i>k>l} dJ_{ikl} \; \exp\left\{-\frac{1}{2p!}J^2_{ikl}2N^{p-1}-J_{ikl}\int dt
[i\hs_i\s_k\s_l+\s_i i\hs_k\s_l+\s_i\s_ki\hs_l]\right\}= \nn \\
&=&\exp\left\{\int\frac{dt dt'}{4N^{p-1}}[p(i\hs\cdot i\hs)(\s \cdot \s)^{p-1}+p(p-1)(i\hs\cdot\s)(\s\cdot i\hs)(\s\cdot\s)^{p-2}]
\right\}
\eea
where we used the notation,
\be
\s\cdot\s\equiv\sum_{i=1}^N \s_i(t)\s_i(t') 
\ee
As we anticipated, in the calculation appeared a coupling among different times, through the overlap of the configuration
at time $t$ and $t'$. In complete analogy with the static case, we therefore introduce as an order parameter the {\it dynamical overlap}
\cite{sompolinsky82,crisanti93}, and get,
\bea
&& \overline{\exp(i\hat\s\cdot {\cal L}_J)} = \nn \\ 
&=&\int DQ\ \d\left(NQ_1-\sum_ki\hs_k(t)i\hs_k(t')\right)
            \d\left(NQ_2-\sum_k\s_k(t)\s_k(t')\right)\cdot\ \nn \\ 
&\cdot&\d\left(NQ_3-\sum_k i\hs_k(t)\s_k(t')\right)
\d\left(NQ_4-\sum_k \s_k(t)i\hs_k(t')\right)\cdot \nn \\
&\cdot& \exp\left\{\frac{pN}4\int dt dt'[Q_1(t,t') Q_2(t,t')^{p-1}+(p-1)Q_3(t,t')Q_4(t,t')Q_2(t,t')^{p-2}]\right\}
\nn
\eea
It is clear the  similarity between the overlap matrix 
$Q_{ab}=\sum_k \s_k^a\s_k^b/N$ in the static approach, and 
$Q_2(t,t')=\sum_k \s_k(t)\s_k(t')/N$: in the first case
we have a static overlap between configurations belonging to 
different replicas, in the second case we have a dynamic overlap between 
configurations at different times. 
It is not only $Q_2$ that has a physical meaning. From their definitions and from 
the discussion above, we see
that $Q_3$ and $Q_4$ are both a response functions, $\langle\s\hat \s\rangle$, 
with their time arguments exchanged. 
Finally, it is possible to argue that the (hard to interpret) order parameter $Q_1=\langle \hat \sigma\hat \sigma\rangle$ 
must be zero \cite{sompolinsky82}. Summarizing,
\be
\left\{
\begin{array}{l}
{Q}_1(t,t')=0 \\
{Q}_2(t,t')=C(t,t')\\
{Q}_3(t,t')=R(t',t) \\
{Q}_4(t,t')=R(t,t')
\end{array}
\right.
\ee
We now give an exponential representations of the $\d$-functions. For example, 
\be
\d\left(NQ_2(t,t')-\sum_k\s_k(t) \s_k(t')\right)= 
\int D\l_2 \exp\left[ iN\int dt dt' \, \left(\l_2 Q_2- \l_2\s\cdot\s\right)\right]
\ee
and we use the saddle-point method to compute the integral. 
By setting to zero the derivatives with respect to all the $Q$'s, we get the
equations,
\be
\left\{
\begin{array}{l}
i\l_1=\frac p4Q_2^{p-1} \nn \\
i\l_2=\frac p4 (p-1)Q_1 Q_2^{p-2}+\frac p4(p-1)(p-2)Q_3Q_4Q_2^{p-3}\equiv 0 \nn \\ 
i\l_3=\frac p4 (p-1)Q_4 Q_2^{p-2} \nn \\ 
i\l_4=\frac p4 (p-1)Q_3 Q_2^{p-2} \\
\end{array}
\right.
\ee
The product $Q_3Q_4$ is zero because of causality: if $t>t'$, then $R(t',t)=0$ and {\it vice versa}.
In order to write the effective Langevin equation we have to recognize what are the new coefficients
of $\hs \hs$ and of $\hs \s$. From the definition of the $\l$'s we have the following  new term in the
generating functional,
\be
\Delta = \sum_k \int dt\; dt' \ \left\{ -\frac{p}{4} C(t,t')^{p-1} \hs_k(t)\hs_k(t') 
- \frac12p(p-1) R(t,t')C(t,t')^{p-2}i\hs_k(t)\s_k(t') \right\} \ ,
\ee
Note that at this point the sites in the action of the generating functional are all decoupled. Therefore we can write
an effective Langevin equation for a scalar degree of freedom $\sigma$ \cite{kirkpatrick87},
\be
\p_t\s(t)=-\mu(t)\s(t)+\frac12p(p-1)\int dt''R(t,t'')C(t,t'')^{p-2}\s(t'')+\xi(t)
\ee
with
\be
\la \xi(t)\xi(t')\ra=2T\d(t-t')+\frac p2 C(t,t')^{p-1}
\ee
Note that the average over the disorder did not generate terms $\s\s$, which we would not
know how to interpret. The effective Langevin equation does not contain the disorder, and
it is uncoupled in the sites. However, it is more complicated than the original one, 
since the noise is no longer $\d$-correlated in time, and we have an explicit memory term
at the r.h.s., that is  a non-local kernel which couples the external time $t$ with all the earlier
times $t''<t$.

\subsection{Equations for the correlation and the response}

We now want to use the effective Langevin equation to write some self-consistent equations
for the correlation and the response function. In order to do this we have to introduce some
useful formal relations. The first one is already known,
\be
R(t,t')=\frac{\p\la x(t)\ra}{\p h(t')}=\la x(t)\hat{x}(t')\ra
\ee
The second relation is the following,
\bea
&& \la \frac{\p x(t)}{\p \eta(t')}\ra = \nn \\
&& \int D\eta\  \exp\left[-\frac12 \eta D^{-1}\eta\right]\frac{\p}{\p \eta(t')}
\int Dx D\hat{x}\  x(t)\exp[\hat{x}(\p_tx+\p_xH)+\hat{x}\eta] \nn \\
&&=\la x(t) \hat{x}(t')\ra = R(t,t')
\eea
The last relation is a little harder to prove,
\bea
\la x(t)\eta(t')\ra= \nn \\
=\int D\eta Dx D\hat{x} \ \exp\left[-\frac12 \eta D^{-1}\eta\right]  \ x(t)\eta(t')
\exp[\hat{x}(\p_tx+\p_x H)+ \hat{x}\eta+j\eta]|_{j=0}=&& \nn \\
= \int  D\eta Dx D\hat{x}\  \exp\left[\frac12 \eta D^{-1}\eta\right]x(t) \frac{\p}{\p j(t')} 
\exp[\hat{x}(\p_tx+\p_x H)+ \hat{x}\eta+j\eta]|_{j=0}=&& \nn \\
=\frac{\p}{\p j(t')} \int Dx D\hat{x} \ x(t) \exp[-\frac12 \hat x D \hat x+\hat{x}(\p_tx+\p_x H)+jDj+jD\hat{x}+\hat{x}Dj]|_{j=0}&&
\nn \\
= \int Dx D\hat{x} x(t)\int dt''D(t',t'')\hat{x}(t'')\exp[S(x,\hat{x}')]&& \nn \\
=\int dt'' D(t',t'') R(t,t'') &&
\eea
Note that this last equation is only valid for Gaussian noise.
We are now ready to write the two equations for the response and the correlation function. To get the equation for the
response we differentiate the effective Langevin equation with respect to the effective noise and average,
\bea \label{eqr}
&&\frac{\p R(t_1,t_2)}{\p t_1}= \frac{\p}{\p t_1}\la \frac{\d\s(t_1)}{\d \xi(t_2)}\ra= 
\la \frac{\d\dot{\s}(t_1)}{\d \xi(t_2)}\ra= \nn \\
=& -&\mu(t_1)R(t_1,t_2)+\frac12 p(p-1)\int_{t_2}^{t_1} dt'' R(t_1,t'')C^{p-2}(t_1,t'')R(t'',t_2)+\nn \\ &+& \d(t_1,t_2)
\eea
The equation for the correlation
is obtained by multiplying the effective Langevin equation by $\sigma$ and averaging,
\bea \label{eqc}
&&\frac{\p C(t_1,t_2)}{\p t_1}= \frac{\p}{\p t_1}\la \s(t_1)\s(t_2)\ra=\la \dot{\s}(t_1)\s(t_2)\ra = \nn \\
=&-&\mu(t_1)C(t_1,t_2)+\frac12 p (p-1)\int_{-\infty}^{t_1} dt'' R(t_1,t'')C^{p-2}(t_1,t'')C(t'',t_2)+
\nn \\ &+&\la \xi(t_1)\s(t_2)\ra
\eea
where we can use the third one of the  relations above and get,  
\be
\la \xi(t_1)\s(t_2)\ra=\int dt'' D(t_1,t'')R(t_2,t'')=2TR(t_2,t_1)+\frac p2 \int_{-\infty}^{t_2} dt'' R(t_2,t'')C^{p-1}
(t_1,t'')
\ee
Because of  causality, the term $2TR(t_2,t_1)$ is zero if $t_2<t_1$, as we shall assume. 
Finally we have to get rid of the Lagrange multiplier $\mu(t)$. Differentiating the constant $C(t,t)\equiv 1$, 
we obtain $[\p_t C(t,t')+\p_{t'} C(t,t')]_{t,t'=s}=0$, giving the equation \cite{crisanti93},
\be
\mu(t_1)=\frac 12 p^2\int_{-\infty}^{t_1}dt''R(t_1,t'')C^{p-1}(t_1,t'')+T
\ee
These are the exact dynamical equations for the PSM. When they were first derived in \cite{kirkpatrick87} it was immediately
noted that they were formally identical to the approximated equations formulated by Mode Coupling Theory (MCT) for structural
glasses \cite{bengtzelius, leutheusser,goetze,reichmann}. 
This observation is at the heart of the theory for the glass-transition in 
structural glasses inspired by $p$-spin spin-glass models \cite{ktw}. The physics of the PSM has probably something to do with 
structural glasses, at least at the dynamical level, and provided that MCT works well. Moreover, it looks like MC theory
must also work in the PSM, since it gives the same equations ! Let us analyze this last point more in detail.

\subsection{Diagrammatic technique and Mode Coupling approximation}

The dynamical equations (\ref{eqr}) and (\ref{eqc}) can indeed be obtained within the Mode Coupling approximation. 
Within this approach we consider the perturbative expansion of the Langevin equation and write all the 
physical quantities using a diagrammatic representations. For the rest of this section see \cite{bouchaud96,cugliandolo-lh}.
For the sake of simplicity, we consider the case of a single scalar degree of freedom $\phi$, with an energy
\be
H=\frac{\mu(t)}{2}\phi^2+\frac{g}{p!}\phi^p
\ee
and we assume that the dynamics of $\phi$ is described by the Langevin equation
\be \label{dinadiag}
\frac{\p \phi}{\p t}=-\mu(t)\phi-\frac{g}{(p-1)!}\phi^{p-1}+\eta
\ee
with the initial condition $\phi(0)=0$. Note that this Hamiltonian is a scalar version, without disorder,
of the $p$-spin one. The thermal noise $\eta$ is defined as in the previous case. 

\begin{figure} 
\begin{center}
\includegraphics[width=0.8\textwidth]{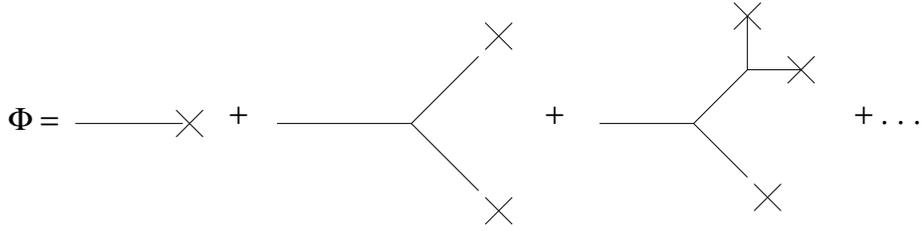}
\end{center}
\caption{Diagrammatic representation of the perturbative solution to equation
\ref{dinadiag}. The various terms of the equation for $\phi(t)$ are
represented: a line stands for the bare propagatore $R_0$, a cross indicates the
noise.  As usual in Feynman diagrams, a vertex stands for a time
convolution.}
\end{figure}

We consider the inverse operator $R_0=[\mu(t)+\frac{\p}{\p t}]^{-1}$, which we use to write the perturbative expansion of $\phi(t)$.
In figure 1 we can see the diagrammatic representation of this expansion in the case $p=3$. In this case
we can write the following equation: 
\be
\phi(t)=R_0 \otimes \eta-\frac{g}{2!} R_0 \otimes \{R_0\otimes \eta \cdot R_0\otimes \eta\}+\ldots
\ee
where $\otimes$ stands for time convolution: $(R_0 \otimes f)(t)=\int_0^t dt'R_0(t,t')f(t')$.
The explicit expression of $R_0$ is
\be
R_0(t,t')=\exp\left[-\int_{t'}^t du\  \mu(u)\right]
\ee
as we can easily see by differentiating $\phi(t)$.
The correlation and response function can be written as,
\be
C(t,t')=\la \phi(t)\phi(t')\ra
\ee
\be
R(t,t')=\la\frac{\p \phi(t)}{\p \eta(t')} \ra = \frac{1}{2T}\la \phi(t)\eta(t')\ra
\ee
\begin{figure} 
\begin{center}
\includegraphics[width=0.8\textwidth]{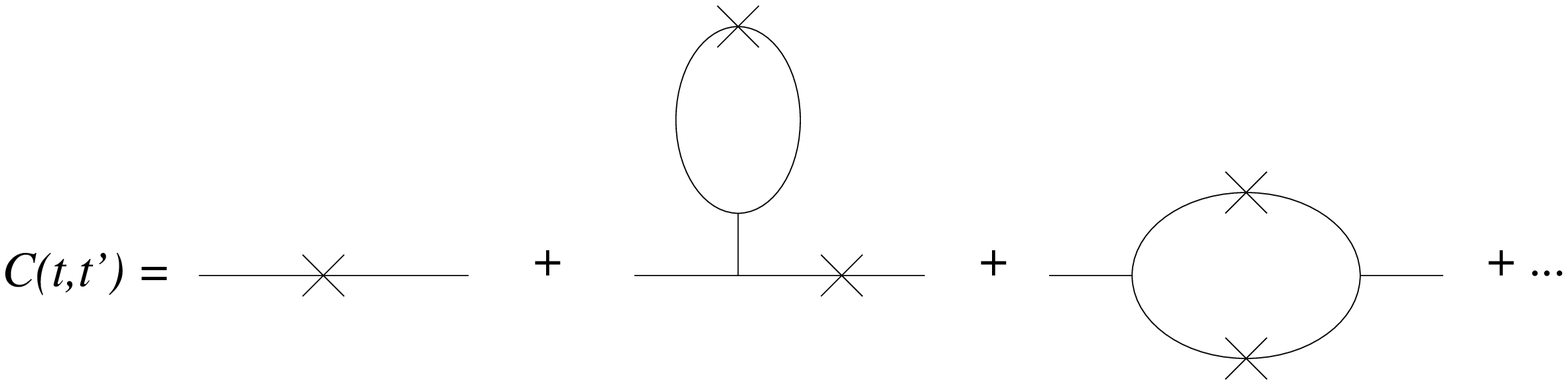}
\includegraphics[width=0.8\textwidth]{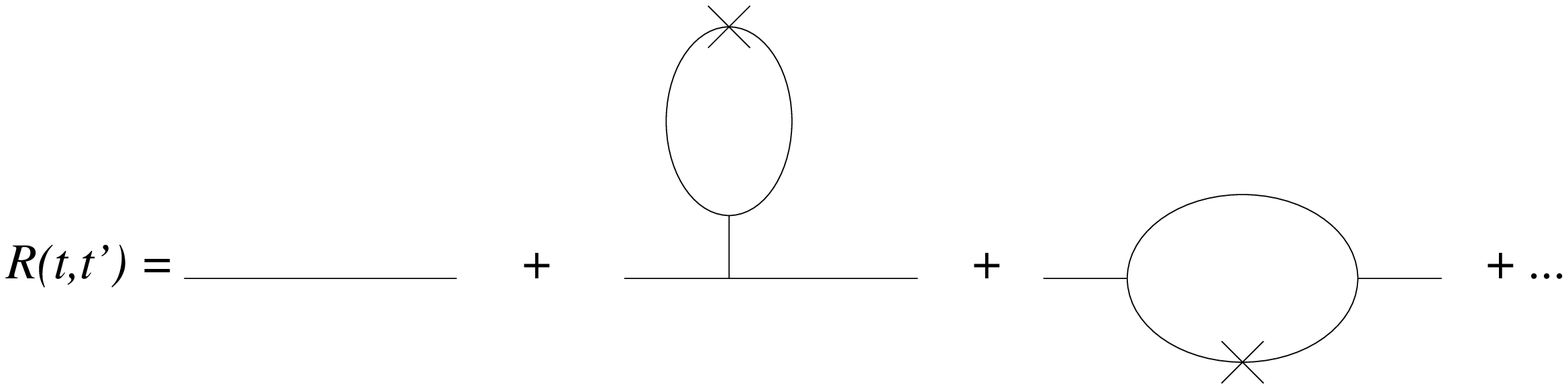}
\end{center}
\caption{
Diagrammatic representations of the perturbative expansions of the
correlation and response function. These diagrams are obtained combining
all the different terms of the diagrams for the perturbative solution.
}
\end{figure}
These functions can be diagrammatically represented, in figure 2 we show the case $p=3$.
In what follows we shall assume that all tadpoles (like the second diagram in figure 2) are already resummed. The contribution
of these diagrams to the self-energy $\Sigma$ that we are going to write is in fact simply a constant: disregarding
tadpoles is equivalent to operate mass renormalization in a usual field theory \cite{fisher-newman,cugliandolo-lh}.
The diagrammatic expansion of $C$ and $R$ can be self-consistently resummed, given the following Dyson equations (fig. 3, upper panel),
\bea
R(t,t')=R_0(t,t')+\int_{t'}^t dt_1 \int_{t'}^{t_1} dt_2 R_0(t,t_1)\Sigma(t_1,t_2)R(t_2,t') \\
C(t,t')=\int_{0}^t dt_1 \int_{0}^{t'} dt_2 R(t,t_1) D(t_1,t_2) R(t',t_2)
\eea 
where the self-energies (or kernels)  $\Sigma(t,t')$ and $D(t,t')$ are, as usual,  the sum of all the amputated connected diagrams.
\begin{figure} 
\begin{center}
\includegraphics[width=0.80\textwidth]{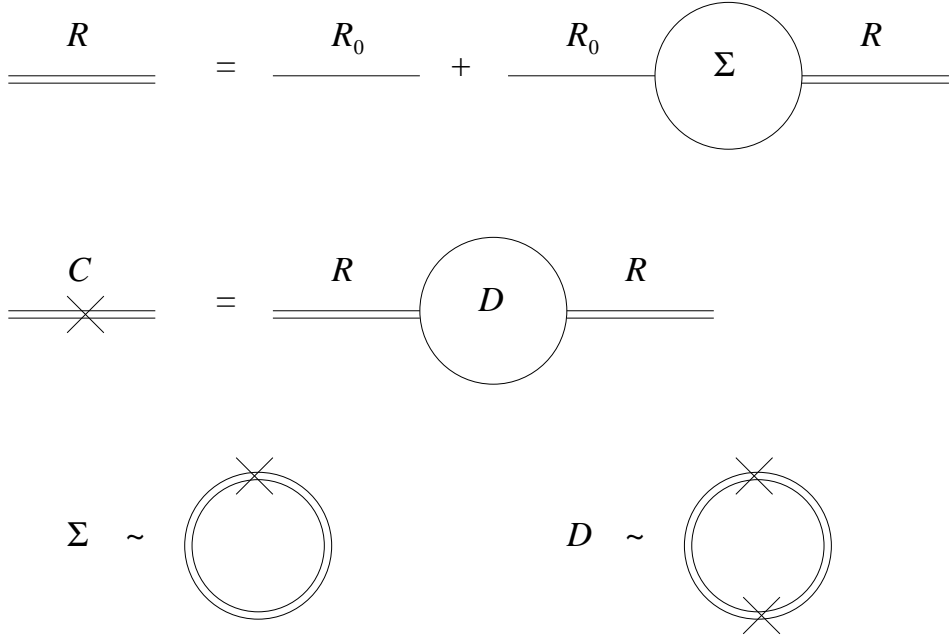}
\end{center}
\caption{The first two lines are the exact Dyson equations for the correlation and the response.
The third line gives the value of the kernels $\Sigma$ and $D$ within the Mode Coupling Approximation.}
\end{figure}
If we multiply by $R_0^{-1}$ we can write the equations in the following way:
\bea
R_0^{-1}\otimes R = I+\Sigma\otimes R \\
R_0^{-1}\otimes C = D\otimes R +\Sigma\otimes C 
\eea
where $I$ is the identity operator. Explicitly we have
\bea
\frac{\p R(t,t')}{\p t} = -\mu(t) R(t,t')+\delta(t-t')+\int_{t'}^t du \Sigma(t,u)R(u,t') \label{tarnu} \\
\frac{\p C(t,t')}{\p t} = -\mu(t) C(t,t')+\int_0^{t'} du D(t,u)R(t',u)+\int_0^{t} du \Sigma(t,u)C(u,t') \nn
\eea
Up to know this was very general, and most importantly exact.
The mode coupling approximation (MCA) consists in approximating the kernels $\Sigma(t,t')$ and
$D(t,t')$: we neglect all the vertex corrections and keep only line corrections, that is we take the values of $\Sigma$ and $G$ at
order $g^2$ and substitute in them the bare response and correlation by their renormalized values, $R_0\to R,\ \  C_0 \to C$. 
In this way we get the following equations:
\bea
\Sigma(t,t') = \frac{g^2}2 C^{p-1}(t,t')R(t,t') \nn \\
D(t,t') = 2T\delta(t-t')+\frac{g^2}6[C(t,t')]^p
\eea
If we now plug this MCA form of $\Sigma$ and $D$ into equations (\ref{tarnu}), it is easy to see that they become 
identical to the  equations we wrote for the $p$-spin model in the previous section
within the generating functional method (we have considered a scalar field, but the same equations can be 
obtained for a vectorial field). This raises an interesting question: how is that the equations obtained with the
Mode Coupling {\it approximation} are identical to those obtained with the {\it exact} generating functional method ?
The answer is that  in a mean-field disordered system, thanks to the scaling with $N$
of the couplings $J$, vertex corrections are sub-leading, and vanish when $N\rightarrow\infty$, while line 
corrections remain finite in the thermodynamic limit. In other words MCA is in fact exact for mean field systems !
 
To understand this fact, 
we can write the Hamiltonian with a vector $\s_i$ (where $i=1,\ldots,N$). The interaction
term for $p=3$ is $g\sum_{i<j<k}J_{ijk}\s_j\s_j\s_k$. In this case the average value of $J^2$ is $1/N^{p-1}$. When 
averaging over the disorder, the behavior in the limit $N\rightarrow\infty$ is different for vertex and line corrections.
In the first case we have (for $p=3$) that in the term $J_{ijk}J_{jlm}J_{mni}J_{kln}$ the average over the noise 
causes the
indices to couple two by two, e.g. $i=l$ and $k=m$. We obtain a factor $(1/N^2)^2$, which we must multiply by $N^4$ having in
the best case
four free indices over which we have to sum. This factor is then of order 1 and vanishes once we normalize the correlation
by $N$. Instead in the line correction the average over the noise causes only two indices to coincide in the best case, giving
a $N$ factor that remains finite also after the normalization \cite{cugliandolo-lh}.

Although a bit boring,  this last section proved an important point: the exact dynamical equations
of the PSM are identical to the MC equations, and in particular to the (approximated) equations that MC theory 
writes for deeply supercooled liquids close to the glass transition. This is one of the main evidences
supporting the idea that the PSM is a sort of mean-field model for structural glasses, and that some of 
 the main  physical concepts valid in the PSM, should be valid in real glasses as well. As long as one believes
 that MC theory describes reasonably well fragile glasses  \cite{angell,kob}, one has to accept that the physics 
 of these systems caught by MC must have something to do with the physics of the PSM.

\subsection{The dynamical transition}

It is now time to solve the dynamical equations of the PSM. In order to do this  
we must make some simplifications on the correlation and response functions.
First, we shall assume that Time Translation Invariance (TTI) holds: 
correlation and response no longer depend independently on the two 
times, but only on their difference. This is true only at {\it equilibrium}, and therefore we are restricting ourselves to 
equilibrium dynamics. The second simplification is in fact a consequence of TTI, 
and is the validity of the Fluctuation-Dissipation Theorem (FDT). These two properties 
can be written as,
\be
\mbox{TTI:}   
\left\{
\begin{array}{l}
C(t_1,t_2)=C(t_1-t_2)\equiv C(\t) \\
R(t_1,t_2)=R(t_1-t_2)\equiv R(\t)
\end{array}
\right.
\quad (\t\equiv t_1-t_2) \\
\ee
\be
\mbox{FDT:}\quad \quad   
R(\t)=-\frac1T \frac {dC(\t)}{d\t} 
\ee
Using these formulas, and a bit of algebra, we can reduce the two coupled equations (\ref{eqr}) and (\ref{eqc})
to a single equations for $C(\tau)$, namely \cite{crisanti93},
\be
\dot{C}(\t)=-TC(\t)-\frac{p}{2T}\int_0^{\t}du \ C^{p-1}(\t-u)\dot{C}(u)
\label{dido}
\ee
A crucial point: in order to perform the integrals we have supposed $C(\infty)=0$, that is we assumed that 
{\it there is no ergodicity breaking}:
after a sufficiently long time the dynamic configuration must be allowed to go as far as possible from the initial 
time configuration. In terms of overlap, this  means that the overlap between $\s(\tau=0)$ and $\s(\tau=\infty)$
must be zero. Recall that we have,
\be
C(\tau)=\frac{1}{N} \sum_k \langle \s_k(\tau)\s_k(0) \rangle
\ee
so that the dynamical correlation function is exactly the average overlap between the two configurations at times $0$
and $\tau$.
Therefore assuming unbroken ergodicity is equivalent to assume that $C(\infty)=0$.
Ergodicity will be verified self-consistently at the end of the calculation.

The first term on the r.h.s. of equation (\ref{dido}) 
comes from the $J$-independent part of the original Langevin equation, i.e. from the 
spherical constraint. The second term, that comes from the 
integration over $J$, is clearly a memory term, for it causes the properties of the system at time $t$ to depend on
all times between $0$ and $t$.

By imposing the physical condition $\dot{C}(\t)\leq 0$ (the average correlation cannot increase with time), 
we obtain from (\ref{dido}) the following relation \cite{notazza},
\be
\underbrace{C^{p-2}(\t)[1-C(\t)]}_{\equiv g(C)}\leq\frac{2T^2}{p} \ .
\ee
This inequality can be easily studied graphically: for $\t=0$ we have $C=1$ and $g(1)=0$, and
for $\t=\infty$ we have $C=0$ and $g(0)=0$. The function $g(C)$ has
thus a maximum between zero and one: let us call $q_d$ the position of this maximum, given by,
\be
q_d = \frac{p-2}{p-1}
\label{qdinamico}
\ee
The r.h.s. of the inequality is a constant larger the larger the temperature.
At very high temperatures the inequality is always satisfied, since $g(q_d) \ll 2T^2/p$. This is the 
paramagnetic phase, which is indeed ergodic. When we lower the temperature, the 
difference between $2T^2/p$ and $g(C)$ gets smaller. From equation (\ref{dido}) we
see that this difference is proportional the time-derivative of $C$: it is large when $C\sim 1$,
that is for short times, it becomes smaller when $C\sim q_d$, and again large for $C\sim 0$, i.e.
for very long times. In other words, when we lower the temperature, we observe the formation of a 
{\it plateau} of the correlation function, with $C(\t)\sim q_d$.

If we lower further the temperature, we arrive at a point where the r.h.s. of the 
inequality touches the curve, i.e. there is a temperature $T_d$ such that $2T_d^2/p = g(q_d)$.
Using (\ref{qdinamico}) and the definition of $g(C)$, we have,
\be
T_d = \sqrt{\frac{p(p-2)^{p-2}}{2(p-1)^{p-1}}}
\ee
At this temperature the correlation function remains stuck  at a plateau $C= q_d$, since $\dot C=0$. Ergodicity 
is therefore broken. We cannot go below $T_d$, since all our assumptions are violated in this phase, and in particular
$C(\infty)\neq 0$.
What we have just proved is that there is a {\it dynamical transition} at $T_d$: the
system passes from the paramagnetic state, to a phase where ergodicity is broken. 
Let us give a physical interpretation of what we have found, in terms of overlap of the 
configuration at time $t$ with the configuration at time $0$.

\begin{description}
\item[$\mathbf{T\gg T_d}$]:
The dynamical overlap (i.e. the correlation function) rapidly decays to zero, such that the configuration
goes as far as it wants in the phase space from its initial position. 
This is the fast equilibrium dynamics in the paramagnetic state.
In this phase relaxation is exponential, and nothing particularly exciting happens.

\item[$\mathbf{T\stackrel{>}{\sim}T_d}$]:
In this phase the dynamics is still ergodic, but something strange happens: for a long time (longer the 
closer to $T_d$ we are) the configuration stays close to its initial value, since $C(\tau)$ has a plateau.
More precisely, the dynamics
explores a ``spot'' of phase space around the initial configuration, of largeness roughly equal to $q_d$, which 
is the value of $C$ at the plateau. This ``spot'' cannot be a true state: if it were, the system would stay 
trapped there forever, while in this case, after a long while, the system drifts away. Eventually 
the overlap goes to zero, consistently with the paramagnetic state. So what is going on ?
Why the system is {\it almost trapped} close to the dynamical transition ? We shall answer these key
questions in the next chapter.

\item[$\mathbf{T\rightarrow T_d^+}$]:
The plateau becomes infinite, the correlation function does not decay anymore, so that the system 
takes an infinite time to equilibrate.  The configuration remains
close to its initial position for an infinite time, and it is clearly trapped by a state of
self-overlap $q_d$. Ergodicity is broken and our dynamical equations break down.

\end{description}
The behaviour of the correlation function we have just described is not a peculiarity of the PSM. In fact, 
it is the typical phenomenology of glassy systems, and in particular of structural glasses. The core aim of
Mode Coupling Theory is indeed to explain this phenomenology in fragile glasses. The interpretation of the 
plateau in finite dimensional glasses is usually given in terms of {\it cage effect}: at low enough temperature,
each particle is surrounded by a cage of nearest-neighbor particles, and it takes a long time  (longer the lower the temperature)
to the particle 
to break this cage and achieve asymptotic relaxation. This interpretation is very nice for structural glasses,
but of course it cannot be applied to the mean-field PSM, where there is no space structure, nor cage. 
However the behaviour of the correlation function is indeed the same. This suggests that the cage effect must
have a deeper interpretation, which must be valid both in finite-dimensional and mean-field systems. We will 
propose such an interpolation in the next chapter.

Let us make a brief summary of the dynamical results. Using the generating functional method we obtained
two exact equations for the correlation and the response, which are formally identical to those obtained 
with the MC approximation. In fact MC is exact for the mean-field PSM. We studied the equations assuming
that the system is at equilibrium and that ergodicity is not broken, in other words we studied the properties of
the (ergodic) paramagnetic state. The correlation function decays to zero, but it develops a plateau as the
temperature lowers. In particular, as $T\to T_d^+$ the plateau diverges and ergodicity is broken in this limit.
Thus, this must be the limit of existence of the paramagnetic phase, and therefore $T_d$ marks a dynamical
transition in the system. The relaxation time (roughly, the time the correlation takes to decay to zero)
diverges at $T_d$. 

A natural question is whether this dynamical transition at $T_d$ coincides with the static one at $T_s$.
The answer is {\it no}. One can easily check that $T_d > T_s$: dynamically, the ergodicity is broken at
a temperature higher than the thermodynamic singularity. Clearly, metastability, combined with the
peculiar features of mean-field, must be responsible for this: metastable states trap the dynamics at $T_d$,
while the thermodynamics cannot detect these states. This is what we suspected when we studied the statics:
at $T_s$ the equilibrium spin-glass states appeared as a secondary peak in the $P(q)$. This peak had
zero weight, but nonzero overlap $q_s$, suggesting that even {\it above} $T_s$ some metastable states existed.
From the dynamics we have a clear evidence that metastable states exist from the simple fact that $T_d>T_s$.
The mean-field nature of the PSM makes the barrier around 
these states infinite, so that the equilibrium states are never reached.

A last important comment: we assumed ergodicity, so we can see ergodicity breaking at $T_d$ as the limit
of validity of our calculation. In fact, what we should do for $T \leq T_d$ is to give up the assumption of TTI and FDT,
and solve the full equations. This is hard, but  can be done (with some suitable approximation) 
\cite{cugliandolo, franz94}. 
What is found in this way is that if the system
starts from a high temperature (random) configuration, it never reaches the static equilibrium energy, and in 
this sense dynamics and statics strongly differ. However, the correlation function behaves differently from what
one may think. The plateau is not infinite below $T_d$, but rather has a length that increases as the earlier
of the two times in $C(t,t')$ increases. This phenomenon is known as {\it aging}, and its description is beyond the scope of
these notes \cite{biroli}. What happens is the following: if we constrain the system to be at {\it equilibrium},
then for  $T\to T_d$ we have divergence of the relaxation time, and thus ergodicity breaking. Therefore we cannot 
study the equilibrium properties of the system for $T\leq T_d$, and we have to give up equilibrium. When this is done,
what we find is a {\it weak} ergodicity breaking below $T_d$, which  is an intrinsically off-equilibrium 
phenomenon \cite{cugliandolo}.

Even though we have an explanation in terms of metastable states of why the two sets of results from statics 
and dynamics differ, it would be nice to have a way to unify the two pictures, and obtain {\it both} results.
This will be achieved in the next chapter.

\section{Complexity}

We have seen that in the PSM  different results are obtained from the static and dynamic approaches. 
Are we able to find a unifying approach, within which it is possible to give
an interpretation of all the results collected until now ? The answer is yes. By now we have
understood that the discrepancy between statics and dynamics is due to the presence
of many metastable states. It is time to directly study these states.

\subsection{What is the TAP free energy ?}

Up to know we have seen the effect of the existence of many pure states only in a indirect way.
In the statics, we had to break the replica symmetry because of ergodicity breaking, but the  
free energy we computed was not the individual free energy of the states, but rather the average free 
energy over all the thermodynamically relevant states. On the other hand, we have seen that
dynamically the ergodicity is broken at a temperature $T_d > T_s$ because of the presence of metastable states.
However, even in that case we could not put our finger on the individual states
trapping the dynamics. To do this we need to introduce the TAP free energy.

Pure states are objects living in an $N$ dimensional phase space: in each state $\alpha$ the local 
magnetizations have a well defined value depending on the site, $m_i^\alpha = \langle \sigma_i\rangle_\alpha$,
and a state is identified by the vector of its magnetizations. Therefore, what we need is a function 
defined on this space, i.e. a function of the local magnetizations $m_i$, whose local minima coincide with the 
pure states of the system. The minimization of such a function must provide a set of equations for
the vector $m_i$, equivalent to the mean field equation for $m$ in the Ising model, $m=\tanh(\beta m)$.
This function is the {\it mean-field free energy}, which is known in the context of spin-glasses as 
Thouless-Anderson-Palmer (TAP) free energy, $f_{\rm{TAP}}(m_1\dots m_N)$ \cite{thouless77}.

It is important to stress that the mean-field, or TAP, free energy is a function of the magnetizations $m_i$ and 
{\it not} of the microscopic degrees of freedom $\sigma_i$. In particular its minima do not necessarily
coincide with the energy minima, that is the minima of the Hamiltonian $H(\sigma_i)$.
In fact, pure states {\it cannot} in general be simply
identified with minima of the energy. The problem is that different energy minima may be separated
by energy barriers which at high temperature are small compared with $k_BT$, and thus belong
to the {\it same} pure state. 
Even though for $T\to 0$ a state essentially collapses onto its lowest energy configuration, 
it is important to keep the two concepts distinct. A pure state $\alpha$, identified by the vector 
$m_1^\alpha\dots m_N^\alpha$, is fundamentally a subcomponent of the Gibbs measure, $\langle\cdot\rangle_\alpha$.
As we have stressed in the first chapter, a pure state enjoys the crucial clustering property, property 
that is meaningless when referred to a simple configuration $\sigma_1\dots\sigma_N$.

The TAP free energy density for the PSM is the following \cite{rieger92, kurchan93, barrat97},
\be
f_{TAP}=-\frac{1}{Np!}\sum_{ikl}J_{ikl}m_i m_k m_l -\frac{1}{2\b}\log(1-q)-\frac{\b}{4}[(p-1)q^p-pq^{p-1}+1]
\label{ftap}
\ee
with,
\be
q=\frac1N\sum_i {m_i}^2 \qquad \mbox{;} \qquad m_i=\la \s_i \ra
\ee
The first term is the energy, the second term is minus the entropy multiplied by the temperature,
and the third one is the so-called reaction term \cite{reaction}.
By setting $m_i=0$ for each $i$, we get $f_{TAP}=-\beta/4$,
the correct result for the paramagnetic state we already met in the statics. The mean-field 
equations are obtained by finding the minima of the TAP free energy, $\partial_{m_i} f_{TAP}=0$
for $i=1\dots N$. However, in order to study these equations, it is 
convenient to change variables \cite{kurchan93}. Let us introduce the new set of variables $\{\sigma_1\dots
\sigma_N; q\}$, defined in the following way,
\be
m_i = \sqrt{q}\;  \sigma_i \quad\quad  \sum_i \sigma_i^2 = N
\ee
These variables $\sigma_i$ (sometimes called angular variable) are formally different from the original
spin degrees of freedom, even though they play a very similar role, and are subject to the same spherical
constraint. In terms of the new variables the TAP free energy becomes,
\be
f_{TAP}(\s_i,q)=\frac1N \ q^{p/2}\ H(\s)+R(q,\b)
\ee
where $H$ is formally the original Hamiltonian (this clarifies why we called $\sigma_i$ the new variables),
and $R$ is the $q$ dependent part in (\ref{ftap}). Now we must minimize this free energy with respect
to the angular variables $\sigma_i$ and the self-overlap $q$, taking into account the spherical constraint.
We have,
\be \label{equazioniTAP}
\left\{
\begin{array}{lll}
\displaystyle\frac{\p f_{TAP}}{\p \s_i}=0 & \rightarrow & 
\left\{
\begin{array}{l}
\frac{\p H(\s)}{\p \s_i}=0 \quad \quad i=1\dots N \\
\sum_i {\s_i}^2=N
\end{array}
\right.
\\
\displaystyle\frac{\p f_{TAP}}{\p q}=0 & \rightarrow & \frac1N \frac p2 q^{\frac p2-1}H(\s)+\frac{\p R}{\p q}=0
\end{array}
\right.
\ee
The first $N$ equations, at fixed values of the random couplings $J$, contain all the complexity of the 
problem: if there are many states, i.e. many solutions of the mean-field equations, it is because of these 
$N$ equations. What is surprising is that in the PSM these equations {\it do not} depend on the temperature !
Moreover, they formally coincide with the minimization equations of the Hamiltonian of the model.
Once we have a solution of these first $N$ equations, call it $\sigma_i^\alpha$, we can compute its zero
temperature energy $E_\alpha = H(\s^\alpha)$, and plug it into the equation for  $q$. This
equation {\it does} depend on $\beta$, so that the self-overlap of a state depends on its zero temperature
energy and on the temperature.

This result is surprising. We said above that in general the minima of the mean-field free energy do 
not coincide with the minima of the Hamiltonian, but we seem to have right here an exception to this rule:
in the PSM minima of the TAP free energy are basically minima of the Hamiltonian. Their positions in 
the phase space does not depend on the temperature, while their self-overlap does. 
In other words, in the PSM  there is a one-to-one mapping between  minima of the
Hamiltonian (the energy) and states, i.e. minima of the free energy. At $T=0$ a state (stable or 
metastable) is just a minimum (absolute or local) of the energy. When $T$ grows energy minima
get dressed up by thermal fluctuations, and become states. So the structure of states of the PSM
is just the structure of minima of the Hamiltonian. 

It is {\it very} important to understand  that this is a peculiar feature of the PSM, due to its homogeneous
nature, and that in general it is not like that. However, in the PSM such a simplification holds. 
If one wants to extend such a simplification to more realistic, finite-dimensional systems
(as structural glasses), it is crucial that {\it the temperature is small enough and times are short}.
Nevertheless, in realistic systems barriers are finite, so that identifying minima of $H$ with states
is in fact conceptually very risky.

In the PSM, thus, the  zero temperature energy density, or {\it bare energy density} $E=H(\s)$ 
of the minima of the Hamiltonian, is the only 
relevant quantity to label states. The self-overlap $q$, the free energy density $f$, and the 
finite temperature energy $\cal E$ thus depend
on the bare energy and on the temperature,
\bea
q=q(E,\b) \nn \\
f=f(E,\b) \\
{\cal E}= {\cal E}(E,\b) = \frac{\p (\b f)}{\p \b}
\eea
Where, of course, ${\cal E}(E,\infty)=E$: the zero-temperature average energy of a state is equal to its bare energy.

\subsection{Definition of complexity (and a problem with the modulus).}

The states of the systems, i.e. the minima of the TAP free energy, have the same structure as the minima
of the Hamiltonian. Therefore, we want to study the structure of minima of $H$, and in particular their 
number. The number of minima $\cal N$ grows exponentially with the size of the system,
\be
{\cal N} \sim e^{N\Sigma}
\ee
The quantity $\Sigma$ is called {\it complexity} in the spin-glass community, and {\it configurational
entropy} in the glass community, where minima of the potential energy are considered.
In order to compute $\cal N$ (and thus $\Sigma$), we have to compute
the number of solutions of the equations,
\be
\frac{\partial H}{\partial\s_i} = 0 \quad \quad i=1\dots N
\ee
By calling $\s_\a$ a solution of these equations, we have,
\be
{\cal N}= \int D\sigma\ \sum_{\a=1}^N \delta(\s-\s_\a)
\ee
By using the standard formula,
\be
\d(\partial H)=\sum_{\a}\frac{\d(\s-\s_{\a})}{|\partial\partial H|}
\ee
we have \cite{bray80},
\be
\N=\int D\s\  \d(\partial H) \ |\partial\partial H|  
\ee
where $\partial\partial H$ is a short-cut for the determinant of the second derivative matrix of $H$ (the Hessian).
Here we have two problems: first, having the modulus in such an equation is algebraically very unpleasant; second, 
in this way we are counting {\it all} stationary points of $H$, not simply minima, but also unstable saddles, which
can hardly be associated to pure states of the system. To solve the first problem we are tempted to disregard 
the modulus, and define,
\be
\hN=\int D\s\  \d(\partial H) \ \partial\partial H 
\ee
However, now we have a very severe problem: this quantity is a topological invariant (the Morse constant),
which has no connections whatsoever with the number of minima \cite{kurchan91}. By disregarding the modulus we are
weighting each stationary point with the sign of its Hessian, such that,
\be
\hN=\int D\s\  \sum_\a \d(\s-\s_\a)\  {\rm sign}(\partial\partial H) = +1 -1 +1 -1 +1 \dots
\ee
The situation seems to be going from bad to worse. However, if we restrict our counting to a {\it fixed} 
energy density level $E$, things improve a lot. What we want to do is to count minima of $H$ which have energy $E$.
To do this we can use the formula,
\be
\N(E)=\int D\s\  \d(\partial H) \ \partial\partial H \ \d(H-E) 
\label{numazzo}
\ee
By restricting ourselves to the level $E$ and by {\it keeping $E$ low enough} we can hope that we are in a
region of the phase space where minima dominate, and thus where the Hessian is positive and the modulus
can in fact be disregarded \cite{cavagna98}. This is certainly true close to the ground state $E_0$. Moreover, the quantity
defined in (\ref{numazzo}) has a further advantage in its very limitation: if we push $E$ high enough to arrive in
a region which is no longer dominated by minima, but by {\it saddles}, we expect to have some instability in 
the calculation due to the change in the sign of the Hessian. Thus, we expect that an instability in our calculation 
will be telling us something relevant about the nature of the stationary points we are counting.

\subsection{The calculation of the complexity.}

To find the stationary points of $H$ with the spherical constraint we can use the Lagrange method. In this way 
we obtain \cite{crisanti95},
\be
-\frac{p}{p!} \sum_{kl} J_{ikl} \s_k\s_l - p\frac1N H(\s) \s_i = 0
\ee
Given that we want to fix the energy density $H(\s)/N=E$, the equations become,
\be
-\frac{p}{p!} \sum_{kl} J_{ikl} \s_k\s_l - p E \s_i = 0
\ee
and thus we have,
\be
\N(E)=\int D\s\  \prod_i\d\left( -\frac{p}{p!} \sum_{kl} J_{ikl} \s_k\s_l - p E \s_i\right) \ 
\det\left(-\frac{p(p-1)}{p!} \sum_{kl} J_{ikl} \s_l - p E \d_{ik}\right)
\label{num}
\ee
with the complexity given by,
\be
\Sigma(E) = \lim_{N\to\infty}\ \frac1N \log \ \N(E)
\ee
In order to average over $J$ we have to understand whether the self-averaging quantity is $\N$ or $\Sigma$.
In general it is the complexity, since extensive quantities are self-averaging, while exponentials are not 
(in the statics we had to average $F$ rather than $Z$).
However, in the PSM we have a further simplification: in absence of external magnetic field we have that
\be
\ov{\log \N}=\log\ov{\N}
\label{pongu}
\ee
and thus we can simply average the number, which is much simpler than averaging the logarithm of it.
Equation (\ref{pongu}) holds because the PSM is a 1RSB system at the static level. It can be proved
\cite{alessia} that  if the static overlap matrix of {\it configurations} is $k$RSB, the corresponding
overlap matrix of {\it magnetizations} is $(k-1)$RSB. For the PSM this implies that the complexity 
calculation is $0$RSB and can therefore be performed at an annealed level.
We give the usual exponential representation of the $\delta$-function,
\be
\prod_i \d(X_i) = \int \frac{D\mu}{(2\pi)^N} \ \exp\left(i\sum_{i=1}^N \mu_i X_i\right)
\ee
On the other hand, for the determinant we can use an integral representation in terms of Grassmann variables (fermions) \cite{zinn},
\be
\det A_{ik} = \int D\bar\psi\,D\psi \ \exp\left(\sum_{ik=1}^N \bar\psi_i A_{ik}\psi_k\right)
\ee
where $\bar\psi$ and $\psi$ are anti-commuting $N$-dimensional Grassmann vectors,
\be
\{\bar\psi_i,\psi_i\}=0
\ee
Note that we could have used commuting variables to write the determinant, but at the price of introducing
replicas \cite{bray80}. So, putting all together, we have,
\be
\Sigma(E)=\frac1N \log \ov{\N(E)}=\frac1N\log\int D\s \frac{D\mu}{(2\pi)^N} D\bar\psi D\psi
\ \ov{\exp[S(\s,\mu,\bar\psi,\psi)]}
\ee
where the action $S$ is given by,
\be
S(\s,\mu,\bar\psi,\psi)= -ipE\sum_i\mu_i\s_i - i\frac{p}{p!}\sum_{ikl} J_{ikl} \mu_i\s_k\s_l 
-pE\sum_i\bar\psi_i\psi_i - \frac{p(p-1)}{p!}\sum_{ikl} J_{ikl} \bar\psi_i\psi_k\s_l
\ee
Part of this action depend on the disorder and it therefore must be averaged over the couplings $J_{ikl}$. This  is not 
difficult to do, since these are Gaussian integral of the form,
\be
\ov{\exp[S_J]} = \prod_{ikl}\int dJ_{ikl} \exp\left[ - \frac12\ J_{ikl}^2 \ \frac{2N^{p-1}}{p!} - J_{ikl}(\dots)\right]
\ee
we only have to be careful about a few technical details: first, the terms $\mu\s\s$ and $\bar\psi\psi\s$ must
be symmetrized before averaging; second, it can be proved that the mixed commuting-anticommuting terms
obtained from the integrals are zero \cite{cavagna98}, so that we can effectively treat separately the commuting and anticommuting 
parts; third, we have to remember that we are integrating on the surface of a sphere of radius $\sqrt N$, due
to the spherical constraint. The integral of the commuting part can be performed exactly, while for the anticommuting 
part we will have to work a bit more. Once the $J$ integral is performed, we obtain,
\be
\Sigma(E) = \left[ -\frac12\log(p/2) +\frac12 -E^2\right] + \frac1N\log \ I
\ee
The term in square bracket comes from the $J$ integral of the commuting part, while $I$ comes from 
the fermionic part,
\be
I=
\int D\bar\psi D\psi
\exp\left[-\frac{1}{4N}p(p-1)\left(\sum_i\bar\psi_i\psi_i\right)^2 - pE\sum_i\bar\psi_i\psi_i\right]
\ee
To treat this integral we use an inverse Gaussian integration (Hubbard-Stratonovich transformation), and write,
\bea
I= \int D\bar\psi D\psi\int d\omega \ 
\exp\left[-\frac{N\omega^2}{p(p-1)} +(i\omega - pE)\sum_i\bar\psi_i\psi_i\right]= \nn \\
\int d\omega\ \exp\left[N\left(-\frac{N\omega^2}{p(p-1)} +\log (i\omega - pE)\right)\right]\nn \\
\int d\omega\ \exp\left[N G(\omega)\right]
\eea
where we have performed the (diagonal) fermionic integral. The crucial feature of this formula is the factor $N$
in the exponential: for $N\to\infty$ we can use the  saddle-point method and write,
\be
I = \exp[N G(\hat\omega)]
\ee 
where $\hat\omega$ satisfies the saddle-point equation,
\be 
\left.\frac{\partial G(\omega)}{\partial\omega}\right|_{\hat\omega}=0
\ee
It is easy to check  that the saddle-point solution $\hat\omega$ lies on the imaginary axis, and thus
it is convenient to define,
\be
\omega=i z
\ee
such that,
\be
G(z)=\frac{N z^2}{p(p-1)} +\log (z - pE) 
\ee
Finally we can write the complexity of the PSM as,
\be
\Sigma(E)=-\frac12\log(p/2) +\frac12 -E^2 +\frac{N\hat z^2}{p(p-1)} +\log (\hat z - pE) 
\label{sigma}
\ee
where the solution $\hat z$ of the saddle point equation $\partial_z G(\hat z)=0$ is,
\be
\hat z= \frac{p}{2}\left(E+\sqrt{E^2 - \frac{2(p-1)}{p}}\right)
\ee
The second root gives a  sub-leading contribution in the thermodynamic limit \cite{crisanti95}.

\subsection{Threshold energy and saddles.}

From the form of $\hat z$ we clearly see that something weird happens when the absolute value of the energy $E$
becomes too small. In fact, for the complexity to be a well defined  physical quantity we must have $\hat z$ 
real. If we define the {\it threshold energy} as \cite{kurchan93,cugliandolo},
\be
E_{th}= - \sqrt{\frac{2(p-1)}{p}}
\ee
we can write,
\be
\hat z= \frac{p}{2}\left(E+\sqrt{E^2 - E_{th}^2}\right)
\ee
We see that $\hat z$ is real, and thus the complexity physically defined, only for,
\be
E\le E_{th}
\ee
What have we obtained ? If we plot the complexity, we see that it is an increasing function of $E$, with
negative second derivative. The complexity is zero at an energy $E_0$: below this energy the complexity
is negative, and thus the number of states is exponentially small in the thermodynamic limit. The energy $E_0$
corresponds thus to the {\it ground state} of the system, the lowest part in our landscape. 
On the other hand, the complexity grows up to 
$E=E_{th}$ beyond which it is no longer defined, since $\hat z$ takes an imaginary part. Therefore,
the interval $[E_0:E_{th}]$ is the physical band of states of the PSM, and all the states with $E>E_0$ are
{\it metastable}.

A natural question at this point is: 
what happens above the threshold energy ? Why is the complexity no longer defined in 
that regime ? In order to answer these questions, we have to remember that we disregarded the modulus
of the determinant of the Hessian, and that we therefore expected to have some problems if minima were
no longer dominant in the energy regime under consideration. This is exactly what is going on here:
above $E_{th}$ minima are not dominant anymore, but unstable saddles are, so the Hessian gets the
contribution of the negative eigenvalues of saddles. To see this we have to recall that the anticommuting (fermionic)
part of our total integral was basically nothing else that the {\it average determinant of the Hessian},
\be
\Delta= \ov{\det\left(\frac{\partial H}{\partial\s_i\partial\s_k}\right)}
\ee
What we have obtained above can thus be rewritten as,
\be
\Delta=\exp\left(\frac{N\hat z^2}{p(p-1)}\right)\ (\hat z-pE)^N
\ee
We see that as long as $E\le E_{th}$, and thus $\hat z$ is real, $\Delta$ is positive (it is easy to see
that the term $\hat z-pE$ is positive). This means that below the threshold the Hessian is {\it on average} 
positive-defined, and this is the same as saying that {\it on average} minima dominate in this energy regime.
On the other hand, for $E>E_{th}$, we have $\hat z= a+ i\, b$, and if we plug this into $\Delta$ we obtain \cite{cavagna-2.2},
\be
\Delta(E)= (-1)^{k(E)N} \ \exp[Ng(E)]
\ee
where $k(E)$ and $g(E)$ are two not-too-complicated functions of $E$.
In this energy regime, thus, the sign of the  determinant oscillates when $N$ goes to infinity. This is
exactly what we would expect from the determinant of a matrix with $k(E)N$ {\it negative eigenvalues}. In fact,
it is possible to calculate the eigenvalue spectrum of the Hessian and prove that $k(E)$ is exactly the fraction
of negative eigenvalues of the Hessian \cite{cavagna-2.2}.

The physical picture is therefore the following: below the threshold the energy landscape is dominated by
minima, and the Hessian is positive on average. In this phase disregarding  the modulus is harmless, and the
complexity we find in this way is well defined. On the other hand, above the threshold the landscape is
dominated by unstable saddles, and the average determinant gets an oscillating part. Having disregarded the
modulus, we detect this transition as the point where the complexity develops an imaginary contribution. 
However, if we are not too picky, we can define a new physical complexity in this phase, by isolating the
factor $(-1)^{kN}$ and taking the logarithm of the real part $\exp[Ng]$. By doing this we are in fact
computing the complexity of saddles dominating at energy $E>E_{th}$ \cite{cavagna01}.

\subsection{The equation for the self-overlap}

What we have said above about the threshold and saddles may seem a bit exotic. In order to 
check all that, it is sound to consider the remaining equation for the self-overlap $q$.
Once we specify the bare energy $E$ of a minimum, we can work out  
the self-overlap of the associated finite $T$ state.  We expect that a {\it bona fide} pure state,
i.e. a minimum of the TAP free energy, must have a well defined self-overlap, indicating roughly 
the size of the state in the phase space. On the other hand, we definitely do not expect saddles
to have a well defined self-overlap, since saddles are not trapping stationary points, and it is hard
(although perhaps not impossible \cite{single}) to define their size. 

Given a solution with bare energy $E$ of the first $N$ equations, the corresponding equation for $q$ reads,
\be
-\frac p2 q^{\frac p2-1}\;E+\frac{1}{2\beta(1-q)}-\frac\beta4
\left[p(p-1)q^{p-1}-p(p-1)q^{p-2}\right]=0
\label{rustu}
\ee
By introducing the auxiliary variable,
\be
y=\frac1Tq^{\frac{p-2}2}(1-q)
\label{lepore}
\ee
we can rewrite the equation as,
\be
p(p-1)y^2+2pEy+2 =0 
\quad\mbox{giving}\quad y=\frac{-E\pm\sqrt{E^2-E^2_{th}}}{p-1}  
\ee
where $E_{th}$ is the {\it same} threshold energy as we have introduced in the calculation of the complexity.
Given that the self-overlap $q=\sum_i m_i^2/N$ must definitely be real, equation (\ref{rustu}) does
not admit solutions for $E>E_{th}$ and the self-overlap is not physically defined above the threshold. This is indeed what we expected: it
is natural to associate a state to a minimum of the energy when we turn on the temperature, but it is not
natural at all to do the same with a {\it saddle}.

What is the self-overlap at the threshold energy ? For $E=E_{th}$ we have $y^2=E_{th}^2/(p-1)^2$ and
thus from (\ref{lepore}),
\be
q_{th}^{p-2}(1-q_{th})^2=T^2\frac2{p(p-1)} 
\ee
This equation gives the self-overlap of the threshold states as a function of the temperature.
We know that the PSM has a purely dynamic transition at $T_d$, where the correlation function, instead
of decaying to zero, remains trapped for an infinite time at a plateau, $C(\t)\to q_d$, for $\t\to\infty$. Our
interpretation of this phenomenon was that the dynamic configuration remains trapped in a region of
the phase space of size (self-overlap) $q_d$. It is tempting to compare this value $q_d$ with 
the self-overlap of the threshold states at $T_d$: indeed these are the {\it highest} metastable states,
so it is reasonable to expect they are responsible for trapping the  
dynamics at $T_d$. Recalling the definition of $T_d$, we have,
\be
T_d^2=\frac{p(p-2)^{p-2}}{2(p-1)^{p-1}} \quad \Longrightarrow \quad q_{th}^2(1-q_{th})^2=\frac{(p-2)^{p-2}}
{(p-1)^p}
\ee
This equation could easily be solved on a computer. However, we have a good guess for the solution:
if we plug into it $q_d=\frac{p-2}{p-1}$, we see that it is identically satisfied. The important 
conclusion is that,
\be
q_{th}(T_d) = q_d
\ee
This result confirms all our expectations: the dynamical overlap at the transition $T_d$ has an asymptotic 
limit equal to the self-overlap of the threshold states. These states therefore are the ones trapping the
dynamics, and forbidding it to relax to the equilibrium values.

Another way for the $q$ equation to stop having solution, is by increasing the temperature, $T>T^\star(E)$,
at fixed bare energy $E$. This means that, even though minima of the energy do not depend on the
temperature, states, i.e. minima of the free energy, do. When the temperature becomes too large,
the paramagnetic states becomes the only pure ergodic states, even though the energy landscape is
broken up in many basins of the energy minima. This is just one particularly evident demonstration of the
fundamental different between pure states and energy minima.
 
\subsection{Life with many metastable states}

We have computed the complexity $\Sigma(E)$, a function of the bare energy $E$ of the minima, which does not
depend on the temperature. All these minima become states when $T\neq 0$. As we have seen, the bare 
energy $E$ and the temperature $T$ are the only variables we need in order  to compute the other properties
at finite temperature. In particular the free energy density of the states is a function
$f=f(E,T)$. 

We now ask what is the role (if any) of the complexity when computing the equilibrium
properties of the system. A first intuitive answer is that there must be no role at all: after all, 
the complexity is zero for ground (stable) states, which rule the equilibrium 
properties of the system, and it is only different from zero for metastable states, which we expect
to have no influence on equilibrium. In fact, it is not like that. We must remember that the PSM is
a mean-field model, where even metastable states have an infinite lifetime, and contribute as stable 
states in partitioning the phase space in ergodic sub-components. So, metastable states do play a 
role also in determining equilibrium properties. 
We could expect, thus, that what follows is valid only for mean-field systems. Strictly speaking, yes.
However, in real system, like supercooled liquids at low temperatures, many of the following observations
apply as well, provided that we pay great attention to the ``states vs energy minima'' issue.
More precisely, in the temperature regime where $T$ is low enough so that the dynamics is activated,
but high enough so that the system is still ergodic and at equilibrium, the dynamics consists in
vibrations inside a potential energy minimum, with some rare jumps among minima. In this regime, 
which is the one close to the Mode Coupling temperature, a phase space decomposition as the one we are
going to explain below is applicable \cite{gold}.

Let us compute the equilibrium partition function $Z$ of the system,
\be
Z=\int D{\s}\exp[-\b H(\s)]=\sum_{\a}\int_{\s\in\a}D\s \exp[-\b H(\s)]=\sum_{\a} Z_{\a}
\ee
where $Z_\a$ is the partition function restricted to state $\a$ (stable or metastable it may be).
We have,
\be
Z_\a=e^{-\b N f_\a}
\ee
and thus,
\be
Z=\sum_a e^{-\b N f_\a}
\ee
In these formulas the free energy density of state $\a$ is $f_\a=f(E_\a,T)$, where $E_\a$ is the bare
energy of state $\a$. We want to pass from a sum over all states to an integral over all bare energies,
\bea
Z &=& 
\sum_{\a}\int dE \ \d(E-E_\a)\ \exp[-\b N f(E,T)]=
\nn \\ &=&\int dE\ \N(E) \exp[-\b N f(E,T)]= \nn \\
&=& \int dE\ \exp\{-\b N[ f(E,T)- T\Sigma(E)] \}\equiv \nn \\
&\equiv& \int dE\  \exp[-\b N \Phi(E,T)]
\eea
where we have defined,
\be
\Phi(E,T)\equiv f(E,T)- T\Sigma(E)
\ee
and where we have used the very definition of the number of states at energy $E$,
\be
\N(E)= \sum_\a \d(E-E_\a)
\ee
In the equation above we can use  the saddle-point method in the limit $N\to\infty$:
the integral is concentrated on the value $\Eq(T)$ which minimize the exponent.
The {\it total equilibrium free energy density} $F_{eq}(T)$ is therefore given by,
\be
F_{eq}(T)=-\frac{1}{\b N} \log Z= \min_{E}[f(E,T)-T\Sigma(E)]=\Phi(\Eq(T),T)
\ee
with
\be
 \frac{\p \Phi}{\p E}(\Eq(T))=0
\label{nardo}
\ee
From its definition we see that $\Phi$ is clearly a sort of {\it generalized free energy}, 
with $f$ playing the role of the energy, and the complexity playing the role of the entropy,
\be
\begin{array}{ll}
f={\cal E}-TS & \mbox{with}\quad S \quad\mbox{entropy} \\
\Phi=f-T\Sigma & \mbox{with}\quad \Sigma \quad\mbox{complexity} \\
\end{array}
\ee
and putting together these formulas we have,
\be
\Phi = {\cal E} - T (S + \Sigma)
\ee
so the complexity is the extra contribution to the total entropy due to the presence of an exponentially
large number of metastable states.
We recall that  $\cal E$ is the {\it finite temperature} energy density of the states, which is
different from the bare (zero-temperature) energy $E$. In fact, ${\cal E}={\cal E}(E,T)$ and ${\cal E}(E,0)=E$.

From what said above we see that the total equilibrium free energy density is found by minimization
with respect to $E$ 
of the potential $\Phi(E,T)$, in which the complexity plays a major role. The bare energy density 
$\Eq(T)$ obtained minimizing $\Phi$, fixes the equilibrium states of the system. The free energy density of
these equilibrium states will then be $f_{eq}=f(\Eq(T),T)$. What is a bit surprising is 
that,
\be
F_{eq} = f(\Eq,T) - T\Sigma(\Eq) < f(\Eq,T)
\ee
since the complexity is positive. In other words, {\it the free energy density of equilibrium states
is larger than the global equilibrium free energy density}. This funny thing is due to the presence of an
exponentially large number of metastable states: equilibrium is given by an ensemble of 
 states, each one with rather large free energy density $f_{eq}$, but whose {\it collective}
contribution to equilibrium is enhanced by their complexity, which lowers the global free energy $F_{eq}$.

This situation may seem paradoxical: equilibrium is given by a mixture of metastable states, but each of
them is surrounded by infinite free energy barriers, so dynamically the system would not be able to 
exit from anyone of these states~! On the one hand, this is just a particular way of breaking the ergodicity,
which is of course strictly valid only in mean-field. On the other hand, this situation makes much more
sense in finite dimensions, where these metastable states may trap the dynamics for a time sufficiently long
to allow us to define a complexity, but sufficiently short to make the system ergodic. This may indeed be the
situation in structural glasses close to the glass transition \cite{gold,stillinger,sciortino}.

\subsection{Low temperatures, entropy crisis}

The interval of definition of $\Phi(E,T)$ is the same as $\Sigma(E)$, that is $E\in[E_0:E_{th}]$.
Assuming that at a given temperature $T$ the energy $\Eq(T)$ minimizing $\Phi$ lies in this interval, what
happens if we lower the temperature ? Remember that the complexity is an increasing function of $E$,
as of course is $f(E,T )$. When $T$ decreases we favor states with lower free energy and lower
complexity, and therefore $\Eq$ decreases. As a result, it must exist a temperature $T_0$, such that,
\be
\Eq(T_0)= E_0
\ee
and thus,
\be
\Sigma(\Eq(T))=\Sigma(E_0)=0
\ee
Below $T_0$ the bare energy $\Eq$ cannot decrease any further: there are no other states below the
ground states $E_0$. Thus, $\Eq(T)=E_0$ {\it for each temperature} $T\le T_0$. As a result, if we plot
the complexity of equilibrium states $\Sigma(\Eq(T))$ as a function of the temperature, we find
a discontinuity of the first derivative at $T_0$, where the complexity vanishes.

A thermodynamic transition takes place at $T_0$:  below this temperature equilibrium is no longer 
dominated by metastable states, but by the lowest lying states, which have zero complexity and lowest
free energy density. The temperature $T_0$ can be computed by studying numerically equation (\ref{nardo}).
The following result should not be surprising at this point,
\be
T_0 = T_s
\ee
The temperature where equilibrium is given for the first time by the lowest energy states, is equal to 
the static transition temperature. Above $T_0$ the partition function is dominated by an exponentially
large number of states, each with high free energy and thus low statistical weight, such that they 
are not captured by the overlap distribution $P(q)$. At $T_0$ the number of these states becomes sub-exponential
and their weight nonzero, such that the $P(q)$ develops a secondary peak at $q_s\neq 0$.

In supercooled liquids, we can give an interesting interpretation of what is going on.
As we have seen the total entropy is the sum of the entropy $S$ inside each state and the complexity $\Sigma$. But in liquids
the entropy of each energy minimum is (at low enough temperatures) very similar to the entropy of
the crystal $S_{CR}$, while the total entropy is just the entropy $S_{LQ}$ observed in the supercooled 
liquid phase.
Thus we can write,
\be
\Sigma(T) = S_{LQ}(T) - S_{CR}(T)
\ee
i.e. the complexity is the excess entropy of the liquid compared to the crystal. Therefore, the temperature
$T_0$ is the temperature where the entropy of the liquid seems to become equal to the temperature
of the crystal, as first observed by Kauzmann in 1948 \cite{kauzmann}.
This scenario (vanishing complexity at $T_0$) is normally known as {\it entropy crisis}.

Of course in real systems we cannot observe
$T_0$, since it is  far below the dynamical glass transition, where the system falls out
of equilibrium. The possible existence of $T_0$ in real liquids relies on low temperature extrapolations
of high temperature equilibrium data.

\subsection{High temperatures, the threshold}

When we raise the temperature we privilege states with higher free energy and complexity. Also in this
case, thus, we must have a temperature beyond which we exit from our range of definition of the
complexity. Indeed, there is a temperature $T_{th}$, such that,
\be
\Eq(T_{th})= E_{th}
\ee
i.e. the bare energy density of equilibrium states becomes equal to the threshold energy at $T_{th}$.
What happens above $T_{th}$ ? If we close our eyes and insist minimizing the potential $\Phi$, we
see that the system would try to thermalize in an energy regime dominated by unstable saddles, and not
by minima. This fact suggests that the dynamics above $T_{th}$ is no longer trapped by minima, and that
therefore it is ergodic. More precisely, we can argue that while below $T_{th}$
equilibrium is in fact given by  a superposition of metastable states with infinite barriers surrounding them,
{\it above} $T_{th}$ the system enters a phase dominated no longer by minima, but by saddles
\cite{cavagna-3,franz,cavagna01}.
A crucial result, which can easily be proved, and which confirms this scenario, is the following, 
\be
T_{th} = T_d
\ee
The temperature $T_{th}$ associated to the transition from minima to saddle (going up in temperature)
is thus the same as the temperature $T_d$ marking the passage from an ergodic to nonergodic dynamics (going
down in temperature). In the light of this, it becomes clearer the interpretation of the dynamics for
$T > T_d$, but close to $T_d$: the landscape visited by the system in this phase is dominated by unstable saddle
points, which have however a very small number of negative modes, since $T \sim T_d=T_{th}$
implies $E\sim E_{th}$. These objects cannot trap the dynamics for infinite times, but they 
can slow it down. In particular, the finite, but very long plateau of the dynamic correlation function $C(\tau)$, 
can be interpreted as a pseudo-relaxation of the system into a saddle with very few (order one) unstable modes \cite{single}. 
At $T_d$ the bare energy is $E_{th}$ and unstable saddles turn into stable trapping minima. 
The plateau becomes infinite and ergodicity is broken.

The identification of $T_d$ with $T_{th}$, and its resulting interpretation, is a crucial point in the
physics of the PSM. It connects the dynamical and topological properties of the system in a very general 
way and it suggests that even in different systems where a glassy transition occurs, the topological
properties of the underlying energy landscape may be the responsible for the slowing down of the system \cite{zuppa}.
We have seen in the previous chapter that the dynamical equations of the PSM are just the Mode Couping 
equations, strongly suggesting that systems well described by MCT close to the glass transition, as
fragile glasses, may have a dynamical behaviour similar to the PSM. As a consequence, one can try to extend
to fragile glasses the topological approach developed in this chapter, which, as we have seen, is so closely
related to dynamics. The cage effect, which as we have seen cannot explain the plateau in a mean-field model, 
can thus be reinterpreted in general as the effect of quasi-stable saddles probed by the system close to the glass 
transition \cite{single}.

A final remark. Mode coupling theory predicts
a sharp transition at $T_d$, but this cannot be strictly true out of mean-field, where barriers are finite.
In fact, even in fragile glasses, at $T_d$ one just observes a very steep crossover, but not a transition \cite{angell}.
However, it may be that the underlying description of the landscape, in terms of minima-to-saddle transition,
is still valid \cite{cavagna-3}. In this way, a unique topological phenomenon  would be responsible for the dynamical 
transition in the mean-field PSM, and of the sharp crossover in finite-dimensional fragile glasses \cite{kurt,angelani,grigera,wales}.

The complexity has at last unified all our results. The two transition temperatures are nothing else than the 
manifestation, at the static and dynamical level, of the lower and upper edges of the band of metastable states.

\section{Conclusions}

It was long enough, so let us be brief in these conclusions. 
We have seen that the PSM has two transitions. There is a thermodynamic transition at a temperature $T_s$,
where the free energy switches from a paramagnetic state, to many spin-glass states. Moreover, $T_s$  
is also the temperature where  the complexity of equilibrium states vanishes. 
What happens at $T_s$ is a perfect realization of the entropy crisis scenario described by Kauzmann 
for supercooled liquids. 
Below $T_s$ equilibrium is given by a 
non-exponential number of lowest free energy states, which are detected and described by a standard
thermodynamic approach. 
Above $T_s$ an exponentially large number of metastable states dominates the 
partition function, due to their nonzero complexity. In this phase, we have the funny result that the 
free energy density of equilibrium states is larger than the global equilibrium free energy density.
Thermodynamics is totally unaware of these states, and predicts a trivial paramagnetic state, with $P(q)=\delta(q)$.

At higher temperatures we have a purely dynamic transition $T_d > T_s$. 
When we arrive at this temperature coming from above, the dynamics gets trapped by metastable states, and the 
correlation time diverges. The equations describing such dynamical behavior are the same as 
the MCT equations for supercooled liquids. This suggests
that what happens in the PSM at $T_d$ is similar to what happens in real glasses close to the MCT temperature. 
The crucial difference, of course, is that in the PSM there can be no barrier crossing, since barriers
are infinite, while in real glasses activation is present. On the other hand, the fact that standard MCT 
predicts a sharp dynamical transition at $T_d$ seems to suggest that this theory too, as the PSM, does not
account for activated events.

We have finally seen that there is a close relationship between the topological properties of the model and
its dynamical behavior. In particular, the slowing down of the dynamics above but close to $T_d$ is
connected to the presence of saddles, whose instability decreases with decreasing energy. In fact, we have seen that
the threshold energy level $E_{th}$ separating saddles from minima, can be associated to the temperature $T_{th}=T_d$,
marking the passage from ergodicity to ergodicity breaking.
In this context the dynamical transition can be seen as a topological transition. The plateau of the 
dynamical correlation function, which has an interpretation in terms of cage effect in liquids, 
may be reinterpreted as a pseudo-thermalization inside a saddle with a very small number of unstable modes. 

A very final warning. We should never forget that the PSM is a mean-field model, with no spatial structure at all. As a consequence, all
physical modelizations and interpretations coming from the PSM, inevitably have a mean-field flavor. In
particular, this is true for the topological interpretation of the dynamical transition: no fluctuations are
taken into account, not to mention spatial heterogeneities, which may play a very important role. However, the
arguments we gave in terms of phase space and topological concepts have at least the virtue of being simple
and effective. If not pushed too far, they provide a nice tool to understand in a unifying  way the physics 
of glassy systems.


\begin{thebibliography}{99}
\bibitem{notazza}
We have essentially used the fact that in the asymptotic limit $\t\gg 1$, 
where $C(\t)\ll 1$, we can write $\int_0^{\t} du  \ C^{p-1}(\t-u)\dot{C}(u)=C^{p-1}(\t)[C(\t)-1]
+\alpha \dot C(\t)$, where $\alpha$ is a positive term.
\end{thebibliography}
\end{document}